\documentclass[a4paper,11pt]{article}

\pdfoutput=1 % if your are submitting a pdflatex (i.e. if you have
\usepackage{jheppub} % for details on the use of the package, please
\usepackage{amsmath}
\usepackage{graphicx}
\usepackage{dcolumn}
\usepackage{bm}
\usepackage{bbm}
\usepackage{epsfig}
\usepackage{slashed}
\usepackage{amssymb}

\makeatletter

\newcommand{\newc}{\newcommand}
\newc{\renewc}{\renewcommand}

\def\beq{\begin{equation}}
\def\eeq{\end{equation}}
\def\bea{\begin{eqnarray}}
\def\eea{\end{eqnarray}}
\def\bitem{\begin{itemize}}
\def\eitem{\end{itemize}}
\def\ba{\begin{array}}
\def\ea{\end{array}}
\def\bal{\begin{align}}
\def\eal{\end{align}}
\def\bi{\begin{itemize}}
\def\ei{\end{itemize}}
\def\lsim{\mathrel{\rlap{\lower4pt\hbox{\hskip1pt$\sim$}}
    \raise1pt\hbox{$<$}}}         %less than or approx. symbolcoupl
\def\gsim{\mathrel{\rlap{\lower4pt\hbox{\hskip1pt$\sim$}}
    \raise1pt\hbox{$>$}}}
\newc{\ie}{{\it i.e.~}}          \newc{\etal}{{\it et al.~}}
\newc{\eg}{{\it e.g.~}}          \newc{\etc}{{\it etc.~}}
\newc{\cf}{{\it c.f.~}}
\newc{\os}{\mbox{\hspace{4pt}}}
\newc{\us}{\mbox{\hspace{12pt}}}

\renewc{\bar}{\overline}

\newc{\gev}{\,{\rm GeV}}
\newc{\mev}{\,{\rm MeV}}
\newc{\ev}{\,{\rm eV}}
\newc{\kev}{\,{\rm keV}}
\newc{\tev}{\,{\rm TeV}}

\newc{\LM}{\mathcal{L}}
\newc{\SM}{\mathcal{S}}

\newc{\HM}{\mathcal{H}}
\newc{\GM}{\mathcal{G}}
\newc{\OM}{\mathcal{O}}
\newc{\FM}{\mathcal{F}}
\newc{\AM}{\mathcal{A}}
\newc{\BM}{\mathcal{B}}
\newc{\NM}{\mathcal{N}}
\newc{\WM}{\mathcal{W}}
\newc{\ZM}{\mathcal{Z}}

\newc{\Chi}{\mathcal{X}}

\title{Phenomenology of Universal Extra Dimensions with Bulk-Masses and Brane-Localized Terms}
\author[a]{Thomas Flacke,}
\author[b]{Kyoungchul Kong,}
\author[c]{Seong Chan Park}

\affiliation[a]{Department of Physics, Korea Advanced Institute of Science and Technology, \\
335 Gwahak-ro, Yuseong-gu, Daejeon 305-701, Korea}
\affiliation[b]{Department of Physics and Astronomy, University of Kansas, Lawrence, KS 66045 USA}
\affiliation[c]{Department of Physics, Sungkyunkwan University, Suwon 440-746, Korea}

\emailAdd{flacke@kaist.ac.kr}
\emailAdd{kckong@ku.edu}
\emailAdd{s.park@skku.edu}

\abstract{
We present a general model with universal extra dimensions 
in the presence of the bulk fermion masses and boundary localized kinetic terms, 
which are generically allowed by symmetries of five dimensional gauge theory. 
We provide a comprehensive analysis for a general UED model, including Kaluza-Klein mass spectra, 
their interactions with the SM particles, and constraints from LHC, electroweak tests, and dark matter experiments.
Finally we show current bounds on the size of allowed universal bulk mass and universal brane-localized terms. 
}


\begin{document} 
\maketitle
\flushbottom

%%%%%%%%%%%%%%%%%%%%%%%%%%%%%%%%%%%%%%%%%%%%%%%%%%%%%%%
%%%%%%%%%%%%%%%%       Intro                 %%%%%%%%%%%%%%%%%%%%%%%%%%%%%%
%%%%%%%%%%%%%%%%%%%%%%%%%%%%%%%%%%%%%%%%%%%%%%%%%%%%%%%

\section{Introduction}
\label{sec:intro}

Although the idea of the existence of extra dimensions has a long history, going back to early twentieth century, 
it has been brought closer to terascale phenomenology only during the last two decades or so \cite{Antoniadis:1990ew}, especially 
in order to understand weakness of the gravitational interaction, supersymmetry breaking, or the identity of the Higgs (for recent reviews see e.g.\cite{ED review}).
Among other models, especially models with universal extra dimensions (UED) \cite{Appelquist:2000nn}, 
where all the Standard Model (SM) particles are allowed to propagate in the extra dimensions, 
address various theoretical issues such as dark matter \cite{servanttait,Cheng:2002ej}, the number of generations \cite{Dobrescu:2001ae}, and the Yukawa hierarchy problem \cite{Csaki:2010az}.
It provides interesting signals that can be probed at various experiments including lower energy, collider and dark matter experiments. 
UED models serve as a nice framework in comparison with supersymmetric models, since the former predicts new particles whose spins are different from those expected in the latter, 
while showing similar signatures. 
A recent study argues that UED can be interpreted as a low energy effective description of Randall-Sundrum model with two-throats  after integrating out the UV regime in the vicinity of the Planck brane \cite{Csaki:2010az}.

A minimal model of UED (MUED) is constructed based on the same gauge structure of the standard model, ${\rm SU(3)_c\times SU(2)_W \times U(1)_Y}$,  on an orbifold $S^1/\mathbb{Z}_2$ or equivalently an interval  $y \in [-L, L]$. The end points $y =\pm L=\pm \pi R/2$ correspond to the fixed points of the orbifold \cite{Appelquist:2000nn}. The standard model fields, gauge bosons, matter fermions and the Higgs, are all promoted to the five dimensional fields.  Every bulk field is decomposed into a tower of Kaluza-Klein (KK) states, in which the zero mode corresponds to a known particle of the standard model. Notably, the action  is invariant under the reflection about the middle point, $y=0$, and the wave functions for bulk fields are either even or odd under this $\mathbb{Z}_2$ parity operation. The parity is called Kaluza-Klein parity (KK-parity), which guarantees the stability of the lightest KK particle (LKP). The LKP, often the first KK mode of the photon ($\gamma_1$), can be a candidate for a dark matter particle 
\cite{servanttait,Cheng:2002ej,Arrenberg:2008wy,Kong:2005hn,Belanger:2010yx,Kakizaki:2006dz}. 
Notice that in MUED neither vectorlike masses 
\cite{Kong:2010qd,Huang:2012kz,sUED1,sUED2,sUEDWp,arXiv:1111.7250,Chen:2009gz} for fermions nor brane terms  
\cite{BLKTrefs,BLKTAPS,FMP} are included.

In MUED \cite{CMS,Cheng:2002ab,Datta:2010us}, all KK particles at level $n$ have the mass, $m_n \approx n/R$ and are almost degenerate. 
Radiative corrections to the KK masses break the degeneracy and play an important role in its phenomenology \cite{CMS}.
Since the cut off scale ($\Lambda$) is not far from the electroweak scale, the mass spectrum turns out to be a rather compressed, 
resulting in soft decay products at the LHC \cite{Cheng:2002ab}. 
Electroweak symmetry breaking also contributes, but it is less significant. 
Radiative corrections also determine the decay patterns of each KK particle, and the LHC with full design luminosity is expected to probe this model beyond $R^{-1} \gsim $ 1 TeV 
\cite{collbounds,2ndKKrefs}.
On the other hand, cosmological considerations give an upper bound on $R^{-1}$, which sets the mass scale of KK particles.
Taking the KK photon as a dark matter candidate, the MUED model predicts $R^{-1}$ to be around 1 TeV \cite{servanttait,Cheng:2002ej,Arrenberg:2008wy,Kong:2005hn,Belanger:2010yx,Kakizaki:2006dz,Chen:2009gz}.

In general, the five dimensional Lagrangian for UED  is composed of bulk terms (${\cal L}_V$) and boundary localized terms (${\cal L}_{\partial V}$).  Staying far away from the end points of the fifth compact dimension, an observer could not notice the presence of boundary so that the bulk action is required to be invariant under five dimensional Lorentz transformations. On the other hand, at the boundaries only the four dimensional subgroup of the full five dimensional Lorentz invariance is respected. In MUED, the boundary terms are assumed to vanish at a cutoff scale $\Lambda$. Naive dimensional analysis yields $\Lambda \sim 24\pi^3/ g_5^2 \sim 24\pi^2/( g_4^2 R)$ since the gauge interactions become strong at this scale. At lower scales, boundary terms are induced at one-loop level and modify the KK spectrum and couplings. However, taking an effective field theory approach to UED models seriously, all  relevant operators beyond the terms in MUED should be also included at the cutoff scale, as long as they are consistent with the required symmetries. Two obvious extensions without extending the field content of MUED include the following terms:
\begin{itemize}
\item Vectorlike masses for bulk fermions (Split-UED)
\item Boundary localized kinetic terms (non-minimal UED)
\end{itemize}
The former (latter) terms are consistent with 5D (4D) Lorentz symmetry as well as the gauge symmetry. Some phenomenological aspects of each term have been already considered in separate contexts 
\cite{Kong:2010qd,Huang:2012kz,sUED1,sUED2,sUEDWp,arXiv:1111.7250,Chen:2009gz,BLKTrefs,BLKTAPS,FMP} 
but the full analysis has not been provided so far, even though its phenomenology could be quite rich and distinct from each individual case.

In this article, we investigate Universal Extra Dimensions in the presence of both vectorlike fermion masses and brane localized kinetic terms. 
We provide the comprehensive analysis for a general UED model, including  
Kaluza-Klein mass spectra,  their interactions with the SM particles, constraints from LHC, electroweak tests, and  dark matter experiments. 
All quantities are given as functions of two additional parameters: the bulk mass $\mu$ and brane coefficient $r$. 
In general, a larger brane coefficient $r$ gives lower KK masses for vectors and fermions, 
while a larger bulk mass $\mu$ increases KK fermion masses. 
Interestingly, interaction couplings of two level-1 KK modes and one level-0 (110) and one level-2 KK mode and two level-0 (200) behave differently as a function of $\mu$. 
We find that they are rather insensitive to $r$.
In general, one can invoke different parameters for each fermion mass and each brane term, making the flavor structure rich but more complicated.
We propose a generalized UED model with a smaller number of new parameters in addition to $R^{-1}$ and $\Lambda$ in MUED.
Given that the LHC is performing very well and putting on stringent bounds on TeV scale physics, 
this generalization would provide an important set up for further LHC studies with more parameters.

Section \ref{sec:model} is devoted to the model, where we perform KK decomposition and calculate relevant masses and couplings, 
which are necessary for further analysis. 
In Section \ref{sec:pheno}, we examine various constraints on the generalized model considering electroweak precision measurements, resonance searches at the LHC, and the relic abundance of KK photon.
Section \ref{sec:conclusion} is reserved for summary and discussion. 
The Appendices  include more details on the flavor issue, KK decomposition, masses and couplings.

%%%%%%%%%%%%%%%%%%%%%%%%%%%%%%%%%%%%%%%%%%%%%%%%%%%%%%%
%%%%%%%%%%%%%%%%       Model                %%%%%%%%%%%%%%%%%%%%%%%%%%%%%%
%%%%%%%%%%%%%%%%%%%%%%%%%%%%%%%%%%%%%%%%%%%%%%%%%%%%%%%
\section{Universal Extra Dimensions with Bulk Masses and Brane Terms}
\label{sec:ued}
%%%%%%%%%%%%%%%%%%%%%%%%%%%%%%%%%%%%%%%%%%%%%%%%%%%%%%%

\subsection{Model}
\label{sec:model}

The model action in five dimensions is invariant under the gauge symmetry of the standard model $ G_{\rm SM}={\rm SU(3)_c \times SU(2)_w\times U(1)_Y}$ consists of two parts:
\begin{enumerate}
\item The action $S_5$ which is invariant under the five dimensional Lorentz symmetry.
\item The boundary action $S_{bdy}\delta(y-y_i)$ where $y_i=(\pm) L$ denotes the location of the end point. The boundary terms are invariant under the four dimensional sub-symmetry of the full five dimensional Lorentz symmetry.
\end{enumerate}
The fermion field content of the model (with a possible extension with the right handed neutrino for the non-vanishing neutrino mass) is given with their charges under the gauge symmetry as follows, 
\begin{eqnarray}
&&Q = (3,2)_{1/6}\ni Q_L^{(0)} = \binom{U_L^{(0)}}{D_L^{(0)}}, \nonumber \\
&&U =(3,1)_{2/3} \ni U_R^{(0)}, \nonumber \\
&&D =(3,1)_{-1/3} \ni D_R^{(0)}, \\
&&L= (1,2)_{-1/2}\ni L_L^{(0)} = \binom{\nu_L^{(0)}}{e_L^{(0)}}, \nonumber \\
&&E=(1,1)_{-1} \ni e_R^{(0)} , ~~(N =(1,1)_{0} \ni \nu_R^{(0)} ) \, , \nonumber 
\end{eqnarray}
where the superscript $^{(0)}$ denotes the zero mode of the Kaluza-Klein tower of the five dimensional field.
The bulk action is given by 
\beq
S_5=\int d^4x \int_{-L}^L dy  \, \left[ {\cal L}_V+{\cal L}_{\Psi}+{\cal L}_H+{\cal L}_{Yuk}\right] ,
\label{5Daction}
\eeq
where  
\begin{eqnarray}
&&{\cal L}_V= \sum_{\AM}^{G,W,B} -\frac{1}{4} \AM^{MN}\cdot \AM_{MN} \\
&&{\cal L}_{\Psi}= \sum_{\Psi}^{Q,U,D,L,E}i \overline{\Psi} \overleftrightarrow{D}_M \Gamma^M \Psi - M_\Psi \overline{\Psi}\Psi \label{FLag}
\end{eqnarray}
where $\AM$ denotes the gluon ($G$), weak gauge bosons ($W$) and the hypercharge gauge boson ($B$) appearing in the gauge covariant derivatives $D_M = \partial_M +i g^5_s \lambda\cdot G_M
+  i g^5_w  \tau \cdot W_M  +  i g^5_Y Y B_M$, where $g^5_i$s are five dimensional couplings of the SM, and $\lambda$s and $\tau$s are the generators of $SU(3)_c$ and $SU(2)_W$, respectively. 
The gauge group indices are suppressed.
$\overline{\Psi} \overleftrightarrow{D}_M \Psi =\frac{1}{2}\{ \overline{\Psi} (D_M \Psi) - (D_M \overline{\Psi}) \Psi \} $. 
The gamma matrix in five dimensions is $\Gamma^M = (\gamma^\mu, i \gamma_5)$, which satisfies $\{\Gamma^A,\Gamma^B\}=2\eta^{AB}=2{\rm diag}(1,-1,-1,-1,-1)$. 
The bulk mass term is chosen to be odd under the inversion about the middle point ($y=0$) of the extra dimension to keep the Kaluza-Klein parity preserved: $M_\Psi(y) = -M_\Psi(-y)$. 

The five dimensional Lagrangian for the Higgs and Yukawa interactions is 
\bea
{\cal L}_H&=& \left(D_M H\right)^\dagger D^M H -V(H), \\
V(H)&=&- \mu_5^2 |H|^2+\lambda_5 |H|^4\,,\label{SHiggs}\\
{\cal L}_{Yuk}&=& \lambda_5^E\overline{L}HE + \lambda_5^D\overline{Q}HD+ \lambda_5^U\overline{Q}\tilde{H}D+\mbox{h.c.}\,,
\eea
where $\tilde{H}=i\tau_2H^*$.

Now for each term in the bulk action, one can add KK-parity conserving boundary terms, that are allowed by gauge invariance and 4 dimensional Lorentz symmetry:
\beq
S_{bdy}= \int d^4 x \int_{-L}^L d y  \, \left({\cal L}_{\partial V}+{\cal L}_{\partial \Psi}+{\cal L}_{\partial H}+{\cal L}_{\partial Yuk}\right) \left[\delta(y-L)+\delta(y+L)\right],
\eeq
with
\bea
{\cal L}_{\partial{V}} &=& \sum_{\AM}^{G,W,B} -\frac{r_\AM}{4}   \AM_{\mu\nu} \cdot \AM^{\mu\nu}, \\
{\cal L}_{\partial \Psi}&=& \sum_{\Psi=Q,L} i  r_{\Psi} \overline{\Psi}_L D_\mu \gamma^\mu \Psi_L  
+\sum_{\Psi=U,D,E} i  r_{\Psi}\overline{\Psi}_R D_\mu \gamma^\mu \Psi_R ,\\
{\cal L}_{\partial H}&=&r_H \left(D_\mu H\right)^\dagger D^\mu H+ r_\mu \mu_5^2 |H|^2-r_\lambda \lambda_5 |H|^4\,,\label{SHiggsbdy}   \label{eq:boundaryhiggs} \\
{\cal L}_{\partial Yuk}&=&r_{\lambda^E} \lambda_5^E\overline{L}HE+ r_{\lambda^D} \lambda_5^D\overline{Q}HD+ r_{\lambda^U} \lambda_5^U\overline{Q}\tilde{H}D+\mbox{h.c.}\,. \label{eq:boundaryyukawa}
\eea

As shown above, the general KK parity preserving 5D UED model contains a large number of new parameters. 
Beyond the size $L$ of the extra dimension,  and the bulk parameters $g_\mathcal{A}$, $\mu_5$, $\lambda_5$, $\lambda_5^{U,D,E}$ -- which in MUED can be directly expressed in terms of the standard model parameters -- the model includes five fermion bulk masses $M_{Q,U,D,L,E}$, as well as the boundary gauge parameters $r_G,r_W,r_B$, the boundary Higgs parameters $r_H, r_\mu, r_\lambda$, five boundary fermion parameters $r_{Q,U,D,L,E}$, and three boundary Yukawa couplings $r_{\lambda^{U,D,E}}$, amounting to a total of 19 additional parameters. 
Studying the full parameter space is beyond the scope of this article, and we need an ansatz to reduce the number of parameters. 
\begin{itemize}
\item First of all, above 19 parameters already assume absence of possible flavor changing neutral current (FCNC). 
A priory, the fermion bulk masses, the fermion boundary parameters and the boundary Yukawa couplings are matrices in flavor space. However, for generic choices, FCNCs are induced at tree-level (\cf Ref.\cite{Csaki:2010az}) which are strongly constrained by various experiments. As shown explicitly in Appendix \ref{app:flavor}, tree level FCNCs are absent if all $M_\Psi$, $r_\Psi$ and $r_{\lambda^{U,D,E}}$ are chosen flavor-blind, which reduces the number of free parameters in the fermion sector to 13. 

\item Different $r_\lambda$s in Eqs. (\ref{eq:boundaryhiggs})-(\ref{eq:boundaryyukawa}) generate (flavor-conserving) mass mixing terms between the different KK fermion modes from the Yukawa interactions. 
As their effects are negligible due to the Yukawa suppression, we already set them to be equal at this stage.

\item For $r_\mu \neq r_\lambda$, the bulk and boundary vacuum expectation values (VEV) do not coincide, which leads to a $y$-dependent VEV. This is a priori not excluded, but it complicates the KK decomposition in the electroweak sector. For $r_\mu = r_\lambda\neq r_H $, one can do the KK decomposition of the Higgs field from the 5D Higgs kinetic term, but in this  case, the mass terms induced from the Higgs potential are not diagonal in this KK basis. They induce mixing between the Higgs KK modes which requires to re-diagonalize the basis, which is to be done numerically. Even for $r_H  = r_\mu = r_\lambda$, one still gets KK mode mixing the EW sector, unless $r_B = r_W = r_H$. For simplicity, we assume a common EW boundary parameter. 
\item In principle, one can introduce two bulk masses and two boundary terms for QCD and EW sectors: $r_Q=r_U=r_D$ and $r_{L}=r_E$, and  $M_Q=M_U=M_D$ and $M_{L}=M_E$. 
For simplicity, in this article we assume universal parameters. 
\end{itemize}
Summarizing, in what follows, we make the simplifying assumption of a universal boundary parameter $r\equiv r_{Q,U,D,L,E}=r_{G,W,B}=r_{H,\mu,\lambda}=r_{\lambda^{U,D,E}}$ and a universal KK-odd fermion bulk mass $\mu \theta(y)=M_{Q,L}=-M_{U,D,E}$ where $\theta(y)=2H(y)-1$ is the step function where $H(y)$ is the Heaviside theta function. 
Therefore, the remaining free parameters are 
 \begin{eqnarray}
 L = \frac{\pi R}{2} &:& {\rm compactification~ scale}\,  ,\\
r &:& {\rm universal ~boundary ~parameter} \, , \\
\mu &:& {\rm universal ~bulk ~mass} \, . 
 \end{eqnarray}
Generically one would expect, $r \sim L$ and $\mu \sim L^{-1}$ ($\frac{r}{L} \sim \mu L  \sim {\cal O}(1)$), since they are allowed by all symmetries of the model. 
The cutoff scale is also a parameter but as shown in literature, the dependence on the cutoff in masses and couplings is usually logarithmic and leads to subdominant effects due to the low cutoff scale.

%%%%%%%%%%%%%%%%%%%%%%%%%%%%%%%%%%%%%%%%%%%%%%%%%%%%%%%
%%%%%%%%%%%%%%%%%%%%%%%%%%%%%%%%%%%%%%%%%%%%%%%%%%%%%%%
\subsection{Kaluza-Klein Decomposition}
\label{sec:KKdecomp}
%%%%%%%%%%%%%%%%%%%%%%%%%%%%%%%%%%%%%%%%%%%%%%%%%%%%%%%

In this section, we perform the Kaluza-Klein decomposition of the UED model with boundary terms and fermion bulk masses. We apply the following standard procedure.
\begin{enumerate}
\item Derive the 5D equations of motion from the quadratic part of the action Eq.~(\ref{5Daction}). 
We do not include contributions from electroweak symmetry breaking in this step, but treat them as corrections after the KK decomposition.
\item Separate the equations of motion into a $x^\mu$ and a $y$ dependent part. 
\item Determine the wave functions and KK masses from the solutions to the $y$ dependent equation of motion (EOM) with the boundary conditions at $y=\pm L$ dictated by the boundary action. 
\item Determine the overall factor by canonically normalizing the KK mode kinetic terms. 
\end{enumerate}
Here we only summarize the results. The detailed calculation can be found in Appendix~\ref{app:KKdecomp}.  

A fermion  $\Psi$ with a left-handed zero mode (\ie $Q$ and $L$) in the presence of a boundary parameter $r$ and a bulk mass $M_\Psi=\mu\theta(y)$ is decomposed as follows.
 \beq
\Psi (x,y)=\sum_{n=0}^\infty \left(\psi^{(n)}_L(x)f^{\Psi_L}_n(y)+\psi^{(n)}_R(x)f^{\Psi_R}_n(y)\right) \, ,
\label{fdecomp}
\eeq
where the wave functions $f^{\Psi_{L/R}}_n$ are given by
\bea
n=0: && f^{\Psi_L}_0=\NM^{\Psi}_0 e^{\mu |y|},\label{fzeromode}\\
\mbox{odd } n:&&\left\{\begin{array}{l}
f^{\Psi_L}_n=\NM^{\Psi}_n \sin(k_n y) \, , \label{ffplus}\\
f^{\Psi_R}_n=\NM^{\Psi}_n \left(-\frac{k_n}{m_{f_n}}\cos(k_n y)+\frac{\mu}{m_{f_n}}\theta(y) \sin(k_n y)\right) \, ,
\end{array}
\right.\\
\mbox{even } n:&&\left\{\begin{array}{l}
f^{\Psi_L}_n=\NM^{\Psi}_n \left(\frac{k_n}{m_{f_n}} \cos(k_n y)+\frac{\mu}{m_{f_n}} \theta(y)\sin(k_n y)\right) \, ,\\
f^{\Psi_R}_n=\NM^{\Psi}_n \sin(k_n y) \, . \label{fKKmodelast}\\
\end{array}
\right.
\eea
The wave numbers $k_n$ are the solutions of the mass quantization condition
\beq
\begin{array}{ll}
 k_n \cos(k_n L)=(r \left(m_{f_n}\right)^2+\mu)\sin(k_n L)  \,  &  ~~{\rm for~ odd~ n} \, , \\
r k_n \cos(k_n L)=-(1+r \mu)\sin(k_n L) \,  & ~~  {\rm for~ even~ n}\,,
\label{fKKnumb}
\end{array}
\eeq
and the masses $m_{f_n}$ of the KK fermions follow from the wave numbers by
\beq
m_{f_n}=\sqrt{k_n^2+\mu^2},
\label{fKKmass}
\eeq
while the chiral zero mode is massless. The normalizations
\beq
\NM^{\Psi}_n= \left\{
\begin{array}{lr}
\sqrt{  \frac{\mu}{ ( 1 +2 r \,  \mu) \exp \left ( 2 \mu L  \right ) - 1}} &\mbox{for } n=0\, ,\\
\frac{1}{\sqrt{L - \frac{\cos (k_n L) \sin(k_n L) }{k_n} + 2 r \sin^2 ( k_n L) } }&\mbox{for odd } n\, ,\\
 \frac{1}{\sqrt{L - \frac{\cos (k_n L) \sin(k_n L) }{k_n}  } }&\mbox{for even } n\, ,
\end{array}
\right.
\label{Nf}
\eeq
are determined from the modified orthogonality relations
 \bea
\int_{-L}^L dy \, f^{\Psi_L}_m f^{\Psi_L}_n[1+r\left(\delta(y+L)+\delta(y-L)\right]=\delta_{mn},\nonumber\\
\int_{-L}^L dy \, f^{\Psi_R}_m f^{\Psi_R}_n=\delta_{mn}.
\label{fermionsclprd}
\eea
A fermion with a right-handed zero mode (\ie $U,D,E$) yields analogous results when replacing $\mu$ with $-\mu$ (see Appendix \ref{app:rhf} for details).
\footnote{As we defined our bulk mass term as  $\mu \theta(y)=M_{Q.L}=-M_{U,D,E}$, the KK masses of the $SU(2)$-doublet and -singlet fields are equal 
(up to corrections from electroweak symmetry breaking), and in this sense, this choice leads to a ``universal'' bulk mass.} 

As has been pointed out in Ref. \cite{BLKTAPS}, in the absence of a bulk mass term, negative boundary parameters lead to a KK spectrum which, depending on the value of $r$, 
contain ghosts and/or tachyons. In the presence of a bulk mass term, we arrive at the same conclusion (see Appendix \ref{sec:light} for details), and 
therefore demand $r>0$, for which neither ghosts nor tachyons are present.

%\bigskip
The KK reduction of gauge bosons and scalars has been discussed in Ref. \cite{FMP}. The fields are decomposed according to
\bea
\AM _\mu(x,y)&=&\sum_{n=0}^\infty \AM_\mu^{(n)}(x)f^\AM_n(y) \label{gaugeKKdecomp} \, ,\\
H(x,y)&=&\sum_{n=0}^\infty H^{(n)}(x)f^\AM_n(y) \, .
\eea
For a uniform boundary kinetic term as considered in this article, the resulting wave functions  are\footnote{For generic choices of the boundary parameters, the KK decomposition in the electroweak sector is more involved. For a detailed discussion and the general solutions, we refer to Ref. \cite{FMP}.}
\bea
n=0: && f^\AM_0(y)= \frac{1}{ \sqrt{ 2 L ( 1 + \frac{r}{L})}}\label{gaugezm} \, \\
\mbox{odd } n:&& f^\AM_n(y)= \sqrt {  \frac{1}{L+ r \sin^2 (k_n L)} }  \sin(k_n y)\label{gaugeom} \, , \\
\mbox{even } n:&& f^\AM_n(y)= \sqrt {  \frac{1}{L+ r \cos^2 (k_n L)} }  \cos( k_n y)\label{gaugeem} \, ,
\eea
where the wave numbers $k_n$ are determined by 
\bea
 \cot (k_n L) &=  r k_n  \,    &~~{\rm for~ odd~ n,}  \\\nonumber
 \tan (k_n L) &= - r k_n  \,   &~~  {\rm for~ even~ n} \, ,
 \label{gaugeKKmass}
\eea
and the corresponding KK masses are 
\begin{equation}
%m_n^2 = k_n^2 + m_0^2 \, , 
m_{\gamma_n}  = k_n  \, . 
\end{equation}
The wave functions satisfy the orthogonality relation
\beq
\int_{-L}^L dy f^\AM_mf^\AM_n\left[1+r\left(\delta(y+L)+\delta(y-L)\right)\right]=\delta_{mn}.
\label{gaugesclprd}
\eeq
As expected, the masses and wave functions of KK scalars and gauge bosons are identical to the masses and $Z_2$-even fermion wave function solutions in the limit $\mu\rightarrow 0$ 
(up to EWSB effects).

\begin{figure}[t]%[htbp]
\centering
\centerline{ 
\epsfig{file=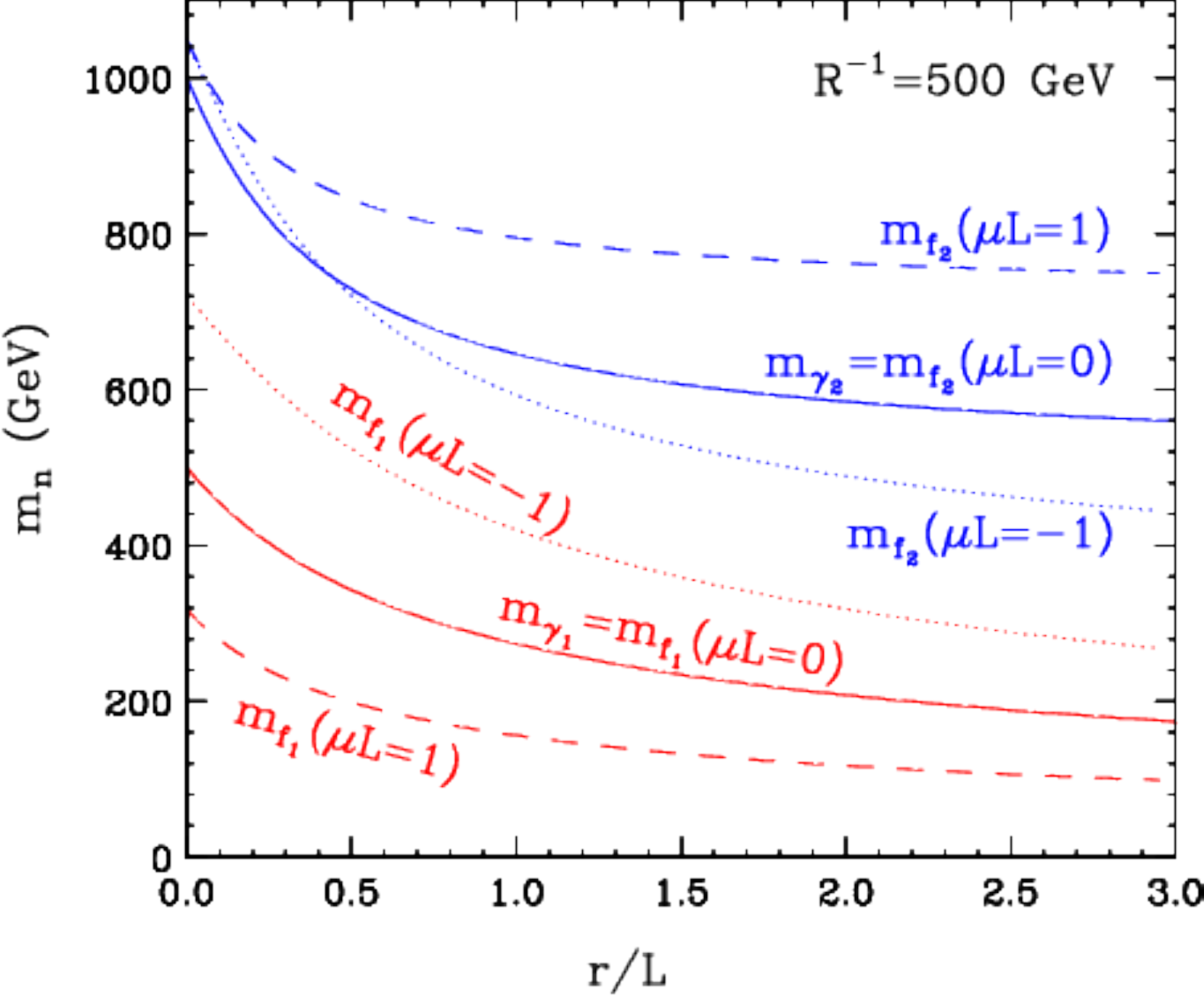, width=0.48\columnwidth} 
%\hspace{-2.8cm}
\epsfig{file=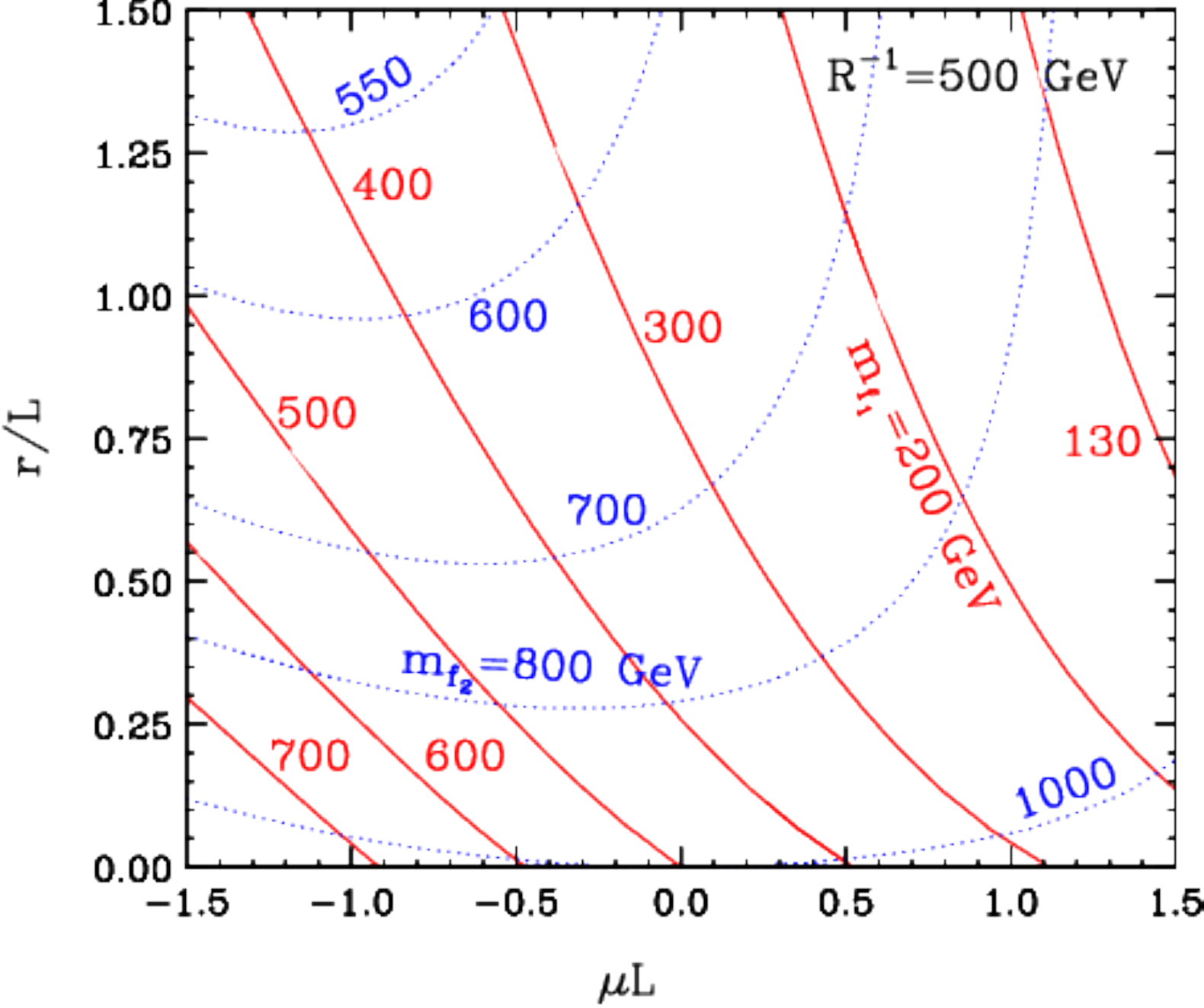, width=0.48\columnwidth}
%\vspace{-4.0cm} 
}
%}
\caption{\sl KK masses at level-1 and level-2 for $R^{-1}=500 \gev$.
}
\label{fig:masses} 
\end{figure}

We close our discussion on the KK decomposition with an illustration of  the dependence of the KK masses on the fermion bulk mass and the boundary parameter shown in Fig.\ref{fig:masses}. 
These masses directly follow from Eqs.~(\ref{fKKnumb}-\ref{fKKmass}). 
In the left panel, we plot masses of the first and second KK fermions, $m_{f_1}$ and $m_{f_2}$, as a function of the dimensionless ratio $r/L$ for different values of the dimensionless parameter  $\mu L$. 
For $\mu L=0$, these masses coincide with the first and second gauge KK mode masses  $m_{\gamma_1}$ and $m_{\gamma_2}$. As can be seen, all KK masses decrease with increasing boundary parameter, while the bulk mass $\mu$ effects the first and second KK modes in opposite ways. While the first KK mode mass is increased for negative $\mu$, the second KK mode mass is decreased in this case -- at least for sufficiently large $r/L$. This non-trivial behavior is a consequence of the different mass quantization conditions Eq.~(\ref{fKKnumb}) for even- and odd-numbered KK modes and  can be seen in more detail in the right panel of Fig.~\ref{fig:masses}, where we plot contours of constant  $m_{f_1}$ and $m_{f_2}$ in the $r/L$ vs. $\mu L$ parameter space. For illustration, we chose a compactification radius  $R^{-1}\equiv2 L / \pi= 500 \gev$ in both figures, but the masses for other compactification radii can easily be deduced, because, as can be seen from Eq.~(\ref{fKKnumb}), the product $m_{f_n} L$ can be expressed as a function of the dimensionless parameters $r/L$ and $\mu L$, only.
In some of the parameter space of Fig.~\ref{fig:masses}, the first KK mode of the fermions is given by hyperbolic solutions rather than trigonometric solutions (\cf  Appendix~\ref{sec:light}, and Fig.~\ref{fig:light} for details). The hyperbolic solutions are taken into account in Fig.~\ref{fig:masses} as well as in our phenomenological studies in Section~\ref{sec:pheno}, but in the phenomenologically viable parameter space identified in Fig.~\ref{fig:colliderdm}, only trigonometric solutions exist.
%

%%%%%%%%%%%%%%%%%%%%%%%%%%%%%%%%%%%%%%%%%%%%%%%%%%%%%%%
\subsection{Couplings between KK modes and SM modes} 
\label{sec:mandcoupl}
%%%%%%%%%%%%%%%%%%%%%%%%%%%%%%%%%%%%%%%%%%%%%%%%%%%%%%%

With the KK decomposition for the fields at hand, we can determine the underlying 5D parameters in terms of the standard model masses and couplings, which in turn fixes all masses of KK modes and their couplings. 

As an example, let us consider the matching of the 4D and 5D gauge couplings. The standard model gauge couplings have to be identified with the couplings of the zero mode fermions to the zero mode gauge bosons which are determined by entering the KK decompositions into the 5D action Eq.~(\ref{5Daction}) and integrating over the extra dimension:
\beq
\begin{split}
S_{eff}&\supset \int d^4x\, i g^5_\AM \overline{\psi}^{(0)} _{L/R}\slashed{\AM}^{(0)}\psi^{(0)}_{L/R} \int^L_{-L} dy f^\AM_0 f^{\Psi_{L/R}}_0f^{\Psi_{L/R}}_0 \left[1+r\left(\delta(y+L)+\delta(y-L)\right)\right]\\
&=\int d^4x\, i \frac{g^5_\AM}{\sqrt{2L(1+r/L)}} \overline{\psi}^{(0)} _{L/R}\slashed{\AM}^{(0)}\psi^{(0)}_{L/R} \, ,
\end{split}
\eeq
implying
\beq
%g^5_\AM=g_\AM\sqrt{2L(1+r/L)},
g^5_\AM=g_\AM\sqrt{  2 L \left (  1+ \frac{r}{L} \right )   } \, ,
\label{4D5Dgauge}
\eeq 
where $g_\AM$ denotes the standard model strong, weak, or hypercharge  gauge coupling. Analogously one finds 
\bea
\mu_5                        &=&  \mu_H  \, , \\ 
\lambda_5                &=&  \lambda_H \left (     2 L \left (  1+ \frac{r}{L} \right )     \right ) \, , \\ 
\lambda_5^{U,D,E }&=& \lambda^{U,D,E}\sqrt{  2 L \left (  1+ \frac{r}{L} \right )   } \, .
\eea
These simple, analytic results hold for our simplifying choice of universal bulk masses and boundary parameters. 
For more general choice, the matching can be performed via the same procedure, but as the $r$ parameters in the orthogonality relations Eqs.(\ref{fermionsclprd}) and (\ref{gaugesclprd}) 
for different fields do not coincide, the matching can in general only be performed numerically (see Ref. \cite{FMP} for an example of non-universal boundary terms in the electroweak sector).

With the 5D parameters determined, the masses and couplings of all KK modes are fixed. 
Beyond the dominant contribution in Eq. (\ref{fKKmass}) and Eq. (\ref{gaugeKKmass}) to the KK mode masses, electroweak symmetry breaking yields an additional contribution to the electroweak gauge boson and fermion KK mode masses. For KK gauge bosons, the relevant term arising from the 5D action in Eq.~(\ref{5Daction}) reads
\bea
S_{eff}&\supset &\int d^4x \left\{\sum_{m,n} \left(-\frac{(g^5_Y)^2 v^2_5}{8}B^{(m)}_\mu B^{(n)\mu}-\frac{g^5_Y g_w^5 v^2_5}{4}W^{3(m)}_\mu B^{(n)\mu}-\frac{(g^5_w)^2 v^2_5}{8}W^{3(m)}_\mu W^{3(n)\mu} \right.\right.\nonumber\\
&&\left.\left.-\frac{(g^5_w)^2 v^2_5}{8}W^{+(m)}_\mu W^{-(n)\mu}
\right)\times
\int_{-L}^L dy f^\AM_mf^\AM_n \left[1+r\left(\delta(y+L)+\delta(y-L)\right)\right]\right\} \nonumber\\
&=& \int d^4x \left\{\sum_n\left( -\frac{m_Z^2}{2} Z^{(n)}_\mu Z^{(n)\mu}-m^2_W W^{+(n)}_\mu W^{-(n)\mu} \right)\right\},
\eea
where in the last step, we used the orthogonality relation Eq.~(\ref{gaugesclprd}), and diagonalized the mass matrix in the neutral sector, which yields the mass eigenstates 
\beq
\left(\begin{array}{c}
A^{(n)}\\
Z^{(n)}
\end{array}\right)=
\left(\begin{array}{cc}
\cos \theta_W  & \sin \theta_W \\
-\sin \theta_W  & \cos \theta_W 
\end{array}\right)
\left(\begin{array}{c}
B^{(n)}\\
W^{3(n)}
\end{array}\right),
\label{gaugemeb}
\eeq
where $\theta_W$ is the standard model Weinberg angle. 
Note that inclusion of radiative correction reduces  the Weinberg angle for KK states \cite{CMS}.
Together with the mass contributions from the KK decomposition, the full masses of the gauge boson KK modes are given by 
\beq
m_{\AM_n}=\sqrt{k_n^2+m_{\AM_0}^2},
\label{gaugemass}
\eeq
where here, $\AM$ denotes the gluon, photon, $Z$-, and $W$-boson, and $m_{\AM_0}$ is the mass of the respective standard model particle. Similarly to the gauge sector, the Yukawa interactions yield additional mass contributions to the KK fermions beyond $m_{f_n}$ from Eq.~(\ref{fKKmass}). The resulting fermion KK masses are given by
\beq
m_{\Psi_n}=\sqrt{m_{f_n}^2+m_{\Psi}^2},
\eeq
where $m_\Psi$ denotes the respective standard model quark and lepton masses.\footnote{The details about the Yukawa contribution to the KK fermion masses and the relation between the gauge- and the mass eigenbasis for KK quarks and leptons can be found in Appendix \ref{app:masses}.}

The couplings of Kaluza-Klein mode particles are determined from overlap integrals of the corresponding wave functions. KK parity guarantees the absence of any KK parity violating interactions. Furthermore, the orthogonality relations guarantee that several couplings are absent (for example KK number violating couplings of fermions to a zero mode gauge boson) or equal to the analogous standard model couplings (for example the coupling between a zero mode gauge boson and two $n$-mode fermions or the coupling between two Higgs zero modes and two $n$-mode gauge bosons). 

\begin{figure}[t]%[htbp]
\centering
\centerline{ 
\epsfig{file=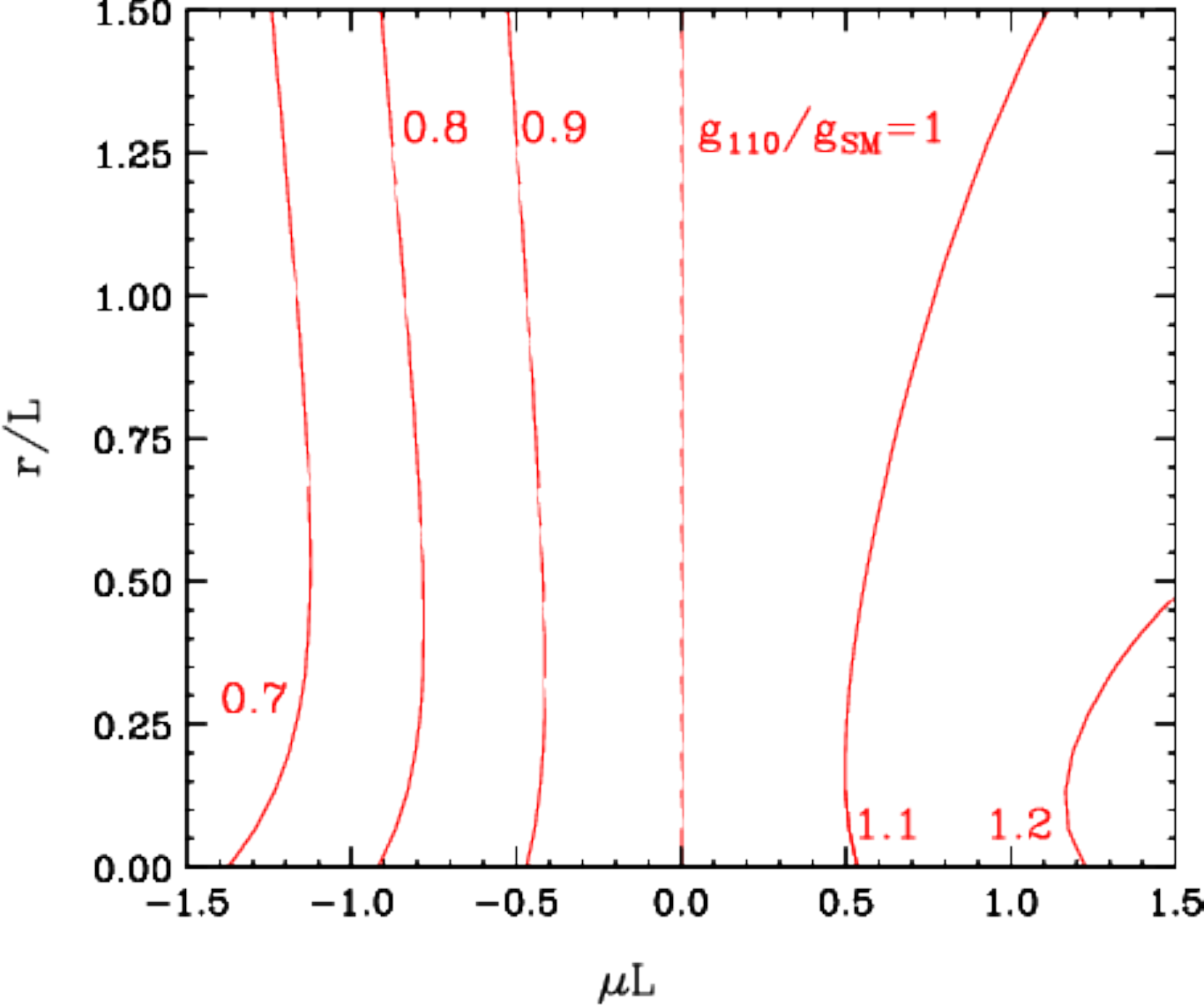, width=0.48\columnwidth} 
\hspace{0.1cm}
\epsfig{file=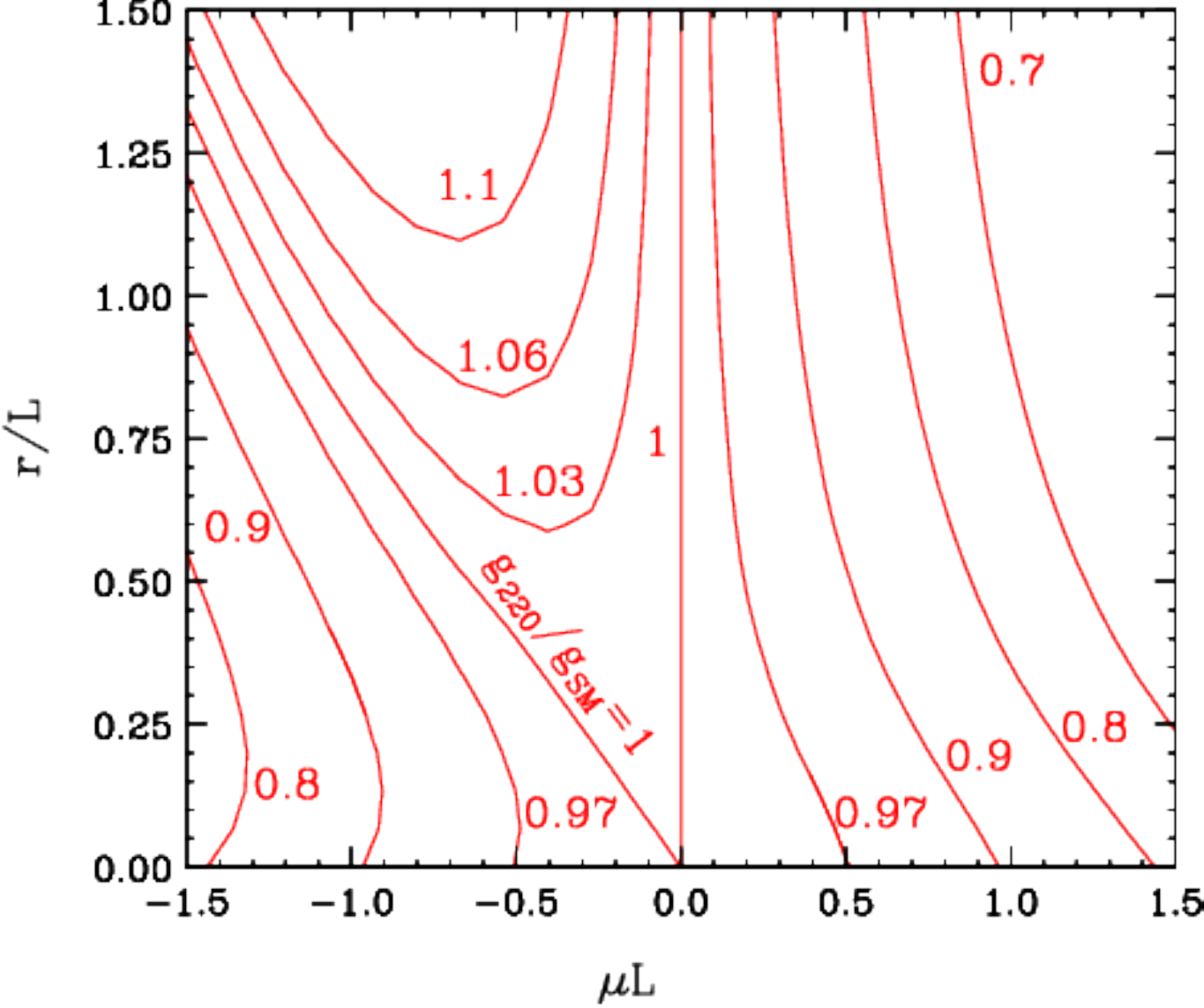, width=0.48\columnwidth}
\vspace{0.3cm}
}
\centerline{ 
\epsfig{file=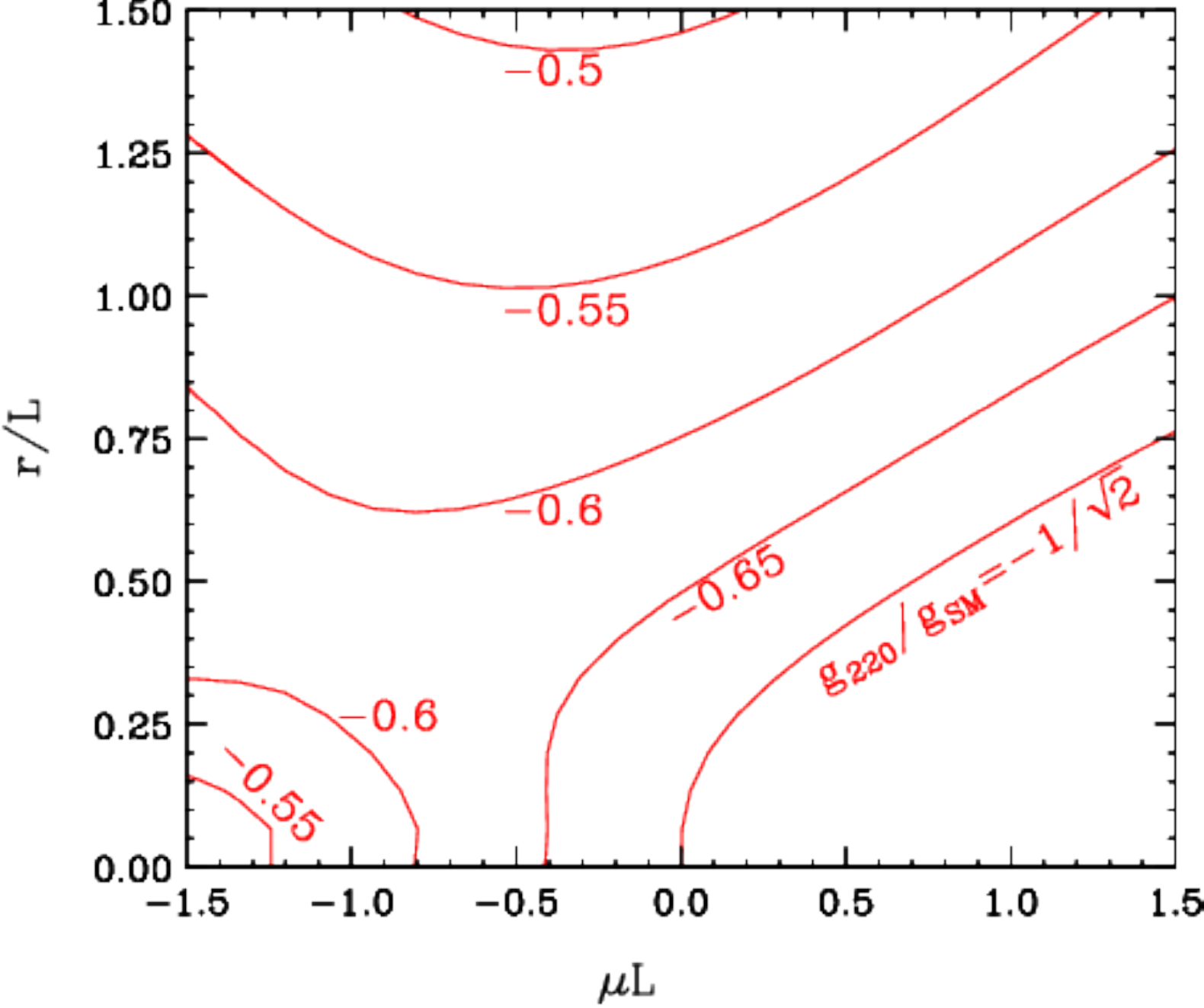, width=0.48\columnwidth} 
\hspace{0.1cm}
\epsfig{file=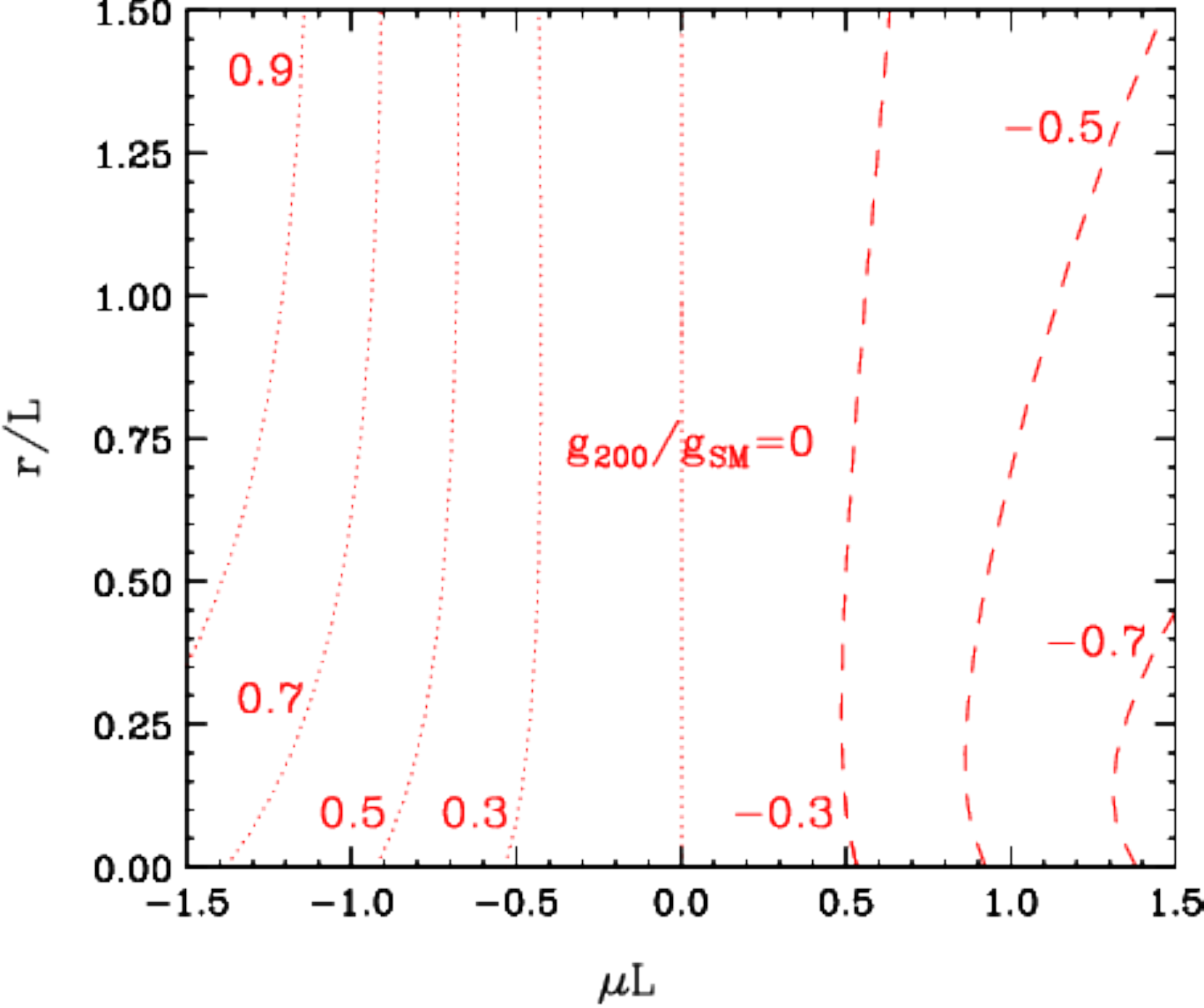, width=0.48\columnwidth}
%\vspace{-3.8cm}
 } 
\caption{\sl Modified KK couplings: $V_1f_1 f_0$ (top-left), $V_2f_2 f_0$ (top-right), $V_2 f_1 f_1$ (bottom-left), and $V_2 f_0 f_0$ (bottom-right).
\label{fig:couplings} }
\end{figure}

Other couplings are modified, compared to minimal UED. The KK number preserving couplings 
\bea
g^\AM_{110}&=& g^5_\AM \int dy \, [1+r \left(\delta(y+L)+\delta(y-L)\right]  f_{1}^\AM f^{\Psi_L}_1 f^{\Psi_L}_0\equiv g^\AM \FM_{110} \, ,
\label{F110def}
\eea
between a zero mode fermion and a one-mode fermion and gauge boson play the dominant role in dark matter annihilation as well as for the production of KK particle pairs and their cascade decays into the LKP at LHC. 
For MUED (at tree level), these couplings are equal to the corresponding standard model couplings. As can be seen in the left panel of Fig.~\ref{fig:couplings}, which shows the ratio $g^\AM_{110}/g_\AM$ as a function of $\mu L$ and $r/L$, the couplings remain equal to the standard model value if no fermion bulk mass term is present (and under the assumption of equal fermion and gauge boundary kinetic terms). This directly follows from the orthogonality relations Eq.~(\ref{gaugesclprd}) and  the fact that for $\mu=0$, the wave functions of the KK-fermions and gauge bosons coincide.  For general $\mu$, however, we observe $\mathcal{O}(1)$ deviations from the standard model couplings. In the usually discussed MUED model, deviations from the standard model couplings are one-loop suppressed. 

The KK number conserving couplings
\bea
g^\AM_{220}&=& g^5_\AM \int dy \, [1+r \left(\delta(y+L)+\delta(y-L)\right]  f_{2}^\AM f^{\Psi_L}_2 f^{\Psi_L}_0\equiv g^\AM \FM_{220} \, ,
\label{F220def}
\eea
between a 2-mode gauge boson and fermion and a zero mode fermion, and 
\bea
g^\AM_{211}&=& g^5_\AM \int dy \, [1+r \left(\delta(y+L)+\delta(y-L)\right]  f_{2}^\AM f^{\Psi_L}_1 f^{\Psi_L}_1\equiv g^\AM \FM_{211} \, ,
\label{F211def}
\eea
between a two 1-mode fermions and a 2-mode gauge boson are also shown in Fig.~\ref{fig:couplings}. Both contribute potential decay channels of the 2-mode gauge boson.

Apart from KK number conserving couplings, the KK number violating couplings 
\beq
g^\AM_{200}=    g^5_\AM \int dy \, [1+r \left(\delta(y+L)+\delta(y-L)\right]  f_{2}^\AM f^{\Psi_L}_0 f^{\Psi_L}_0\equiv g^\AM \FM_{200} \, ,
\label{F200def}
\eeq 
between two zero mode fermions and a level-2 KK mode gauge boson play an important role for collider phenomenology. 
Via these couplings, 2-mode gauge bosons can be produced as an $s$-channel resonance, which implies  $Z'$-, $W'$- or coloron-like signatures. In MUED, these couplings are only induced at one-loop level and therefore small \cite{CMS}, but still potentially observable at LHC when upgraded to $14 \tev$ \cite{2ndKKrefs}. As can be seen in the right panel of Fig.~\ref{fig:couplings}, for our generalized UED setup, the coupling is absent only for $\mu=0$ -- again due to coinciding fermion and gauge boson wave functions and the orthogonality relations. For generic $\mu$, $g^\AM_{200}$ is of the order of the corresponding standard model coupling. Therefore, resonance searches are amongst the most sensitive tests of generalized UED models. 
We find that dependence on the brane parameter $r$ is weak in $\FM_{110}$ and $\FM_{200}$ and we expect that they may be less constrained by experiments. 
On the other hand, variation of $\FM_{220}$ and $\FM_{211}$ along the $r$ direction is more dramatic.

%%%%%%%%%%%%%%%%%%%%%%%%%%%%%%%%%%%%%%%%%%%%%%%%%%%%%%%
%%%%%%%%%%%%%%%%%%           Phenomenology              %%%%%%%%%%%%%%%%%%%%%%
%%%%%%%%%%%%%%%%%%%%%%%%%%%%%%%%%%%%%%%%%%%%%%%%%%%%%%%

%%%%%%%%%%%%%%%%%%%%%%%%%%%%%%%%%%%%%%%%%%%%%%%%%%%%%%%
\section{Constraints on Generalized UED Models}
\label{sec:pheno}
%%%%%%%%%%%%%%%%%%%%%%%%%%%%%%%%%%%%%%%%%%%%%%%%%%%%%%%

In this section, we consider various constraints on the generalized UED model in the presence of bulk masses and brane localized terms.

%%%%%%%%%%%%%%%%%%%%%%%%%%%%%%%%%%%%%%%%%%%%%%%%%%%%%%%
%%%%%%%%%%%%%%%%%%%%%%%%%%%%%%%%%%%%%%%%%%%%%%%%%%%%%%%
\subsection{Electroweak Precision Measurements}
\label{sec:ew}
%%%%%%%%%%%%%%%%%%%%%%%%%%%%%%%%%%%%%%%%%%%%%%%%%%%%%%%

A strong bound on the MUED model arises from UED contributions to the Peskin-Takeuchi parameters $S$, $T$, and $U$ \cite{ewbounds}, which parameterize the oblique corrections to the electroweak gauge boson propagators \cite{PT1}. In the first part of this section, we study the impact of boundary terms and fermion bulk masses on the $S$, $T$, and $U$ parameters. By comparing the UED contributions to the experimentally determined bounds \cite{ewGfitter} we determine the resulting constraints on the generalized UED parameter space.

Another class of electroweak precision measurements which have been performed off the $Z$-pole at several experiments furthermore lead to stringent bounds on four-fermion interactions  \cite{ewlow,pdg}. 
For MUED, these bounds do not lead to relevant constraints on the model because KK-number violating interactions 
-- which are needed for beyond standard model contributions to four-fermion interactions at low energies -- are only induced at loop level and therefore small. 
As shown in the last section, for a more general UED setup, sizable interactions between zero mode fermions and even-KK mode gauge bosons are present, which, when integrating out the non-zero KK modes, lead to four-fermion interactions. In the second part of this section, we use the bounds on four-Fermi interactions to obtain another constraint on the generalized UED parameter space.

%%%%%%%%%%%%%%%%%%%%%%%%%%%%%%%%%%%%%%%%%%%%%%%%%%%%%%%
\subsubsection{Universal Corrections}
\label{sec:stu}
%%%%%%%%%%%%%%%%%%%%%%%%%%%%%%%%%%%%%%%%%%%%%%%%%%%%%%%

\begin{figure}[t]%[htbp]
\centering
\centerline{ 
\epsfig{file=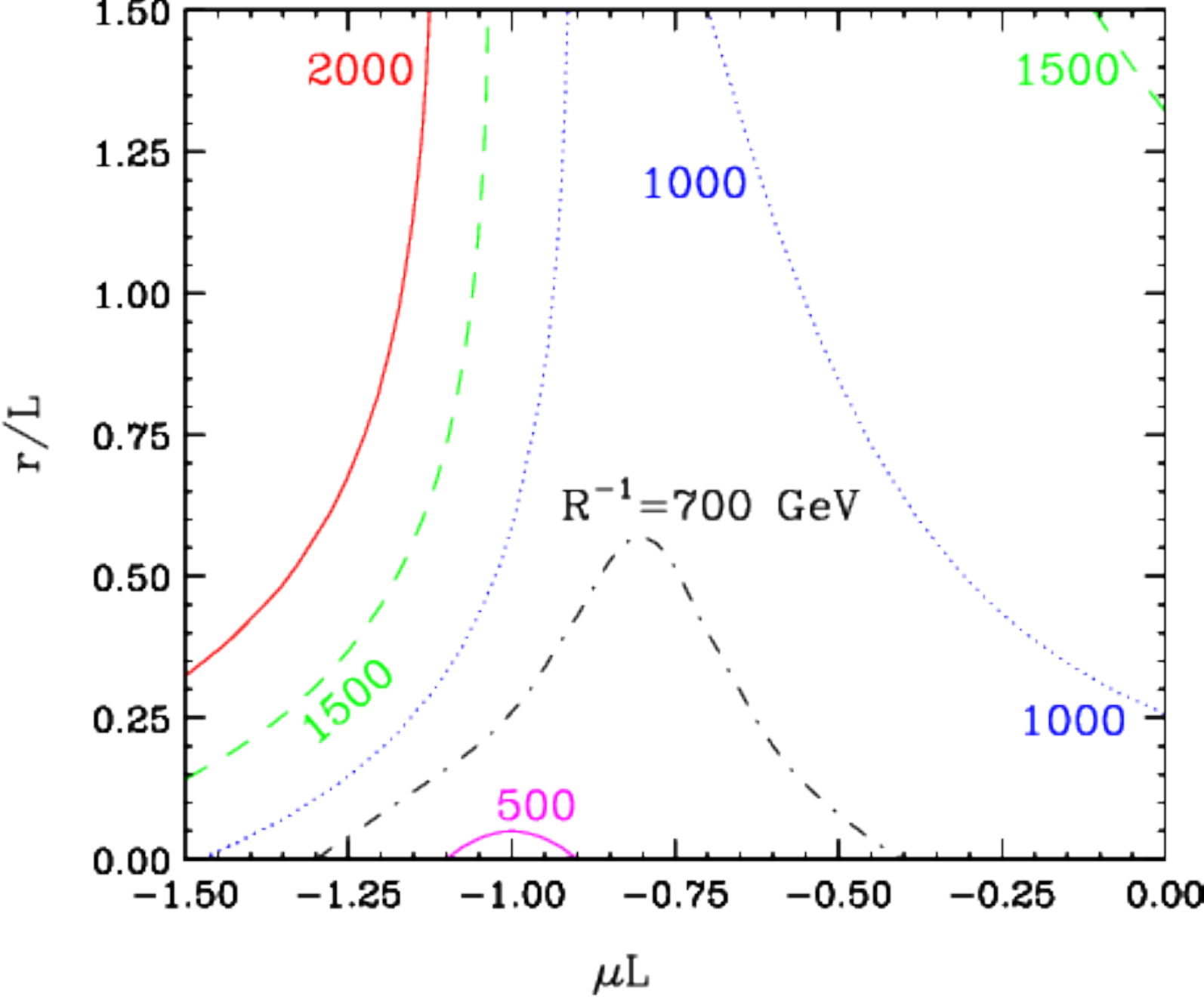, width=0.48\columnwidth} 
\hspace{0.2cm}
\epsfig{file=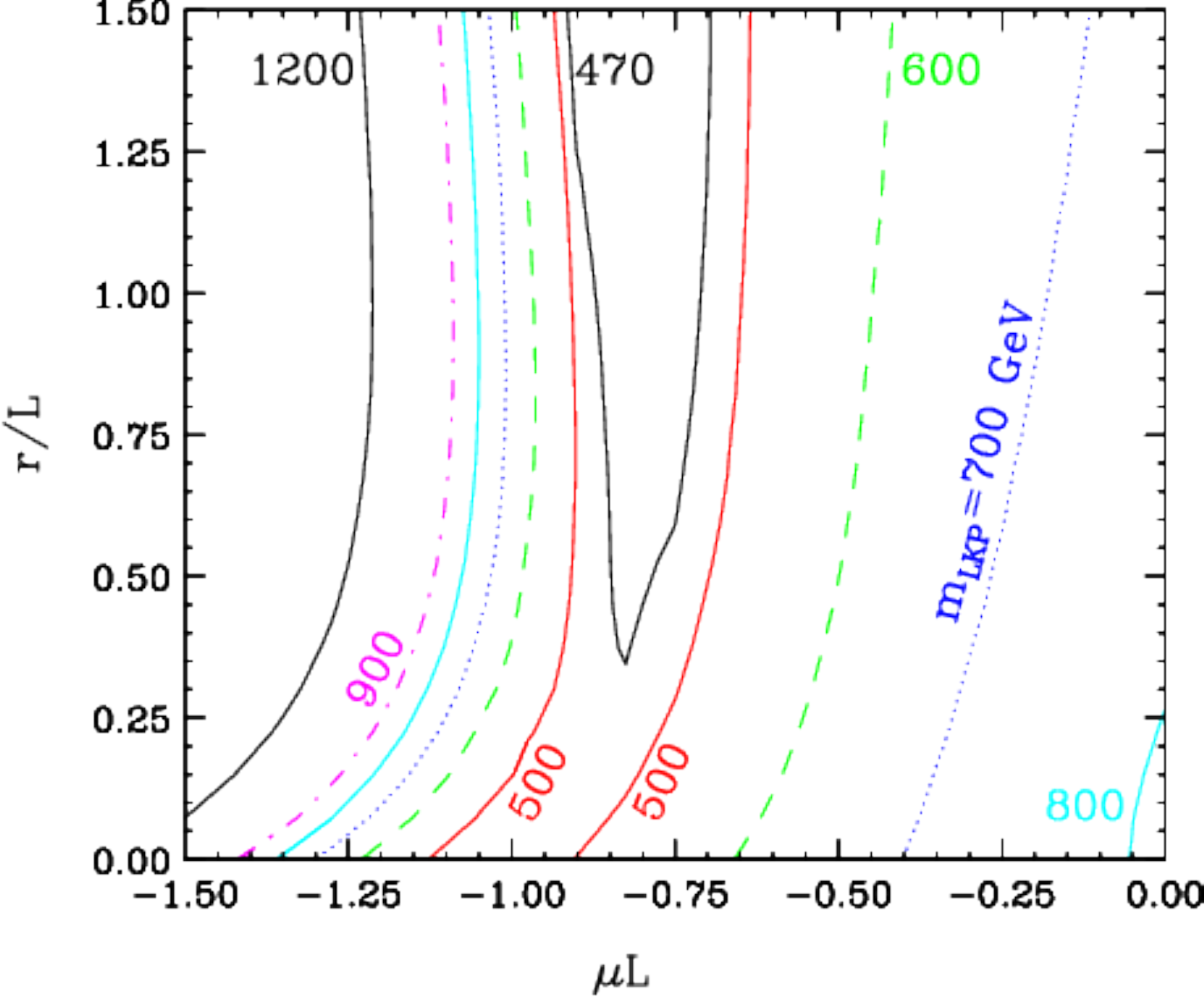, width=0.48\columnwidth} }
\caption{\sl Oblique bounds on $R^{-1}$. Contours represent minimally allowed values of $R^{-1}$.}
\label{fig:oblique} 
\end{figure}

The electroweak corrections in the model under consideration are not oblique in the strict sense because the exchange of even-numbered gauge boson KK modes induce effective four-fermion vertices amongst the standard model fermions. For our choice of common fermion boundary and mass terms, the induced four-Fermi interactions are universal, however. The dominant effect of universal four-fermion operators on electroweak precision observables  arises due to modifications to the Fermi constant $G_f$ which is determined from muon decay, \ie a four-fermion process. The calculation of the Fermi constant yields 
\bea
G_f &=&G^0_f+\delta G_f \, , \\
%\eeq
%with
%\beq
G^0_f &=& \frac{g_{ew}^2}{\sqrt{32}  m_W^2}       ~ \mbox{ and } ~
\delta G_f =  \frac{1}{\sqrt{32}}\sum_n\frac{\left(g^{ew}_{002n}\right)^2}{m_{W^{(2n)}}^2} \, .
\label{Gfcorr}
\eea
$G^0_f$ is simply the contribution from  $W$ zero-mode exchange, while  $\delta G_f$ denotes the sum of the contributions from all non-zero $W$ KK  modes. 
As has been shown in Ref. \cite{CTWP}, such corrections to $G_f$ can be incorporated into the electroweak fit by matching the experimentally determined values of the new physics contributions $S_{NP},T_{NP}$ and $U_{NP}$ to effective parameters. For similar approaches see Refs. \cite{preCTWP}. In our study they are given as follows. 
\bea
S_{eff}&=&S_{UED},\nonumber\\
T_{eff}&=&T_{UED}-\frac{1}{\alpha}\frac{\delta G_f}{G_f} ,\\
U_{eff}&=&U_{UED}+\frac{4 \sin^2\theta_W}{\alpha}\frac{\delta G_f}{G_f}. \nonumber
\label{STUeff}
\eea
The oblique corrections  $S_{UED}$, $T_{UED}$ and $U_{UED}$ at one-loop-level have been studied in Refs.~\cite{ewbounds}. The leading contributions are\footnote{Refs. \cite{ewbounds} calculated their results assuming the MUED tree-level mass spectrum and couplings and expressed their results as a function of $R^{-1}$. The result we give below is expressed in terms of the KK-mode masses and agrees with  Refs.~\cite{ewbounds} in the limit $\mu,r\rightarrow 0$. Note that KK masses depend on $\mu$ and $r$, but the couplings are just given by the standard model couplings. See Section \ref{sec:mandcoupl} for couplings.}
\bea
S_{UED}&=&\frac{4 \sin^2\theta_W}{\alpha}\left[\frac{3g^2_{ew}}{4(4\pi)^2}\left(\frac{2}{9}\sum_n \frac{m^2_t}{m^2_{t^{(n)}}}\right)+\frac{g^2_{ew}}{4(4\pi)^2}\left(\frac{1}{6}\sum_n \frac{m^2_h}{m^2_{h^{(n)}}}\right)\right], \label{Sloop}\\
T_{UED}&=&\frac{1}{\alpha}\left[\frac{3g^2_{ew}}{2(4\pi)^2}\frac{m_t^2}{m_W^2}\left(\frac{2}{3}\sum_n \frac{m^2_t}{m^2_{t^{(n)}}}\right)+\frac{g^2_{ew}\sin^2\theta_W}{(4\pi)^2\cos^2\theta_W}\left(-\frac{5}{12}\sum_n\frac{m^2_h}{m^2_{h^{(n)}}}\right)\right], \label{Tloop}\\
U_{UED}&=&-\frac{4\sin^2\theta_W}{\alpha}\left[\frac{g^2_{ew}\sin^2\theta_W}{(4\pi)^2}\left(\frac{1}{6}\sum_n\frac{m^2_W}{m^2_{W^{(n)}}}-\frac{1}{15}\sum_n\frac{m^2_Wm^2_h}{m^2_{W^{(n)}}m^2_{h^{(n)}}}\right)\right]. \label{Uloop}
\eea
To obtain a bound on the parameter space, we perform a $\chi^2$ fit of the parameters $S_{eff}$, 
$T_{eff}$, $U_{eff}$ from Eq.~(\ref{STUeff}) to the experimental values  given in Ref.~\cite{ewGfitter}:
\beq
S_{NP}=0.03\pm0.10\, ,\, T_{NP}=0.05\pm0.12\, , \, U_{NP}=0.03\pm0.10,
\label{STUexp}
\eeq
for a reference point $m_{h}=126 \gev$ and $m_{t}=173 \gev$ with  correlation coefficients  of $+0.89$ between $S_{NP}$ and $T_{NP}$, and $-0.54$  $(-0.83)$ between $S_{NP}$ and $U_{NP}$ ($T_{NP}$ and $U_{NP}$). 
Fig.~\ref{fig:oblique} shows the implied constraints on the parameter space at 90\% confidence level. 
We represent bounds in terms of the dimensionless parameters $r/L$ and $\mu L$. 
The left panel of Fig. \ref{fig:oblique} shows contours of the minimally allowed compactification scale $R^{-1}$. For $r/L=0$, 
our bounds are consistent with the electroweak constraints on the split-UED model discussed in Refs. \cite{arXiv:1111.7250,Huang:2012kz}.\footnote{The bound on the compactification scale is weakest for $\mu L\sim -0.9$, 
which arises due to the interplay of the tree level and the one loop contributions to $S_{eff}$,  $T_{eff}$ and $U_{eff}$ (see Ref.~\cite{arXiv:1111.7250} for a detailed discussion). 
The constraints we obtain differ slightly from the former studies because we used the updated best-fit values of $S,T$, and $U$ from Ref.~\cite{ewGfitter}.}
For an increasing $r/L$ and a fixed $R^{-1}$, the KK masses of the fermions and gauge bosons decrease. As the UED contributions to the $S,T,U$ parameters scale with $m^{-2}_{KK}$, this implies that the allowed window for $\mu L$ decreases for increasing $r/L$ when $R^{-1}$ is held fixed. Unlike for minimal UED, the compactification scale $R^{-1}$ does not bear a direct physical interpretation in terms of the mass scale of the first KK excitations of the SM fields. In the right panel of Fig.~\ref{fig:oblique}, we therefore in addition present the electroweak bounds in terms contours of the minimally allowed LKP mass in the $r/L$ vs. $\mu L$ parameter space. 
%

%%%%%%%%%%%%%%%%%%%%%%%%%%%%%%%%%%%%%%%%%%%%%%%%%%%%%%%
\subsubsection{Four-Fermi Interactions}
\label{sec:fermi}
%%%%%%%%%%%%%%%%%%%%%%%%%%%%%%%%%%%%%%%%%%%%%%%%%%%%%%%

Beyond the universal electroweak corrections at the $Z$-pole, specific four-Fermi interactions have been strong constraint through measurements at lower energies \cite{ewlow,pdg}. Parameterizing the four-Fermi interactions according to
\beq
\LM_{eff}\supset \sum_{f_1,f_2}\sum_{A,B=L,R}\eta^{s}_{f_1f_2,AB}\frac{4\pi}{(\Lambda^{s}_{f_1,f_2,AB})^2}\bar{f}_{1,A}\gamma^\mu f_{1,A} \bar{f}_{2,B}\gamma_\mu f_{2,B},
\eeq
where $f_{1,2}$ are the contributing fermions and $\eta^{s}_{f_1f_2,AB}=\pm 1$, the strongest bounds on the suppression scales $\Lambda$ can be found in Table~\ref{t:4fermibds}. 
In UED, the four-Fermi operators are induced by the exchange of even-numbered KK-mode gauge bosons which for the $e_Le_Lq_Lq_L$ operators amount to the contributions \cite{Huang:2012kz}
\bea
\LM^{UED}_{eff}&\supset&
4\pi N_c \sum_{n=1}^{\infty}  \left({\cal F}_{2n\, 0 0 } (r/L,\mu  L)\right)^2 
                 \times \left[ \frac{3}{5}\frac{\alpha_1Y_{e_A} Y_{q_B} }{Q^2-M_{B^{(2n)}}^2} +\frac{\alpha_2T^3_{e_A} T^3_{q_B} }{Q^2-M_{W^{3(2n)}}^2}\right]\nonumber\\
 &\approx&
-12 \pi \sum_{n=1}^{\infty}  \left({\cal F}_{2n\, 0 0 } (r/L,\mu  L)\right)^2 
                 \times \left[ \frac{3}{5}\frac{\alpha_1Y_{e_A} Y_{q_B} }{M_{B^{(2n)}}^2} +\frac{\alpha_2T^3_{e_A} T^3_{q_B} }{M_{W^{3(2n)}}^2}\right]                
                    \, ,
\label{eq:eeqq}
\eea
where $\FM_{2n\, 0 0}$ is the overlap integral defined analogously to Eq.~(\ref{F200def}), $Y$'s and $T$'s are the hypercharges and isospins of the corresponding fermions, and in the second line we used  $N_c=3$ and $M_{B^{(2n)}}^2,M_{W^{3(2n)}}^2\gg Q^2$ for the processes from which the experimental bounds originate. The  $U(1)_Y$ and $SU(2)$ fine-structure constants at one-loop level are given by
\begin{eqnarray}
\alpha_1(\mu)=\frac{5}{3}\frac{g_Y^2(\mu)}{4\pi} = \frac{\alpha_1 (m_Z)}{1-\frac{b_1}{4\pi}\alpha_1(m_Z) \log \frac{\mu^2}{m_Z^2}},\\
\alpha_2(\mu)=\frac{g^2_{ew}(\mu)}{4\pi} = \frac{\alpha_2 (m_Z)}{1-\frac{b_2}{4\pi}\alpha_2 (m_Z) \log \frac{\mu^2}{m_Z^2}},
\end{eqnarray}
with $\alpha_1(m_Z)\approx 0.017$,  $\alpha_2(m_Z)\approx 0.034$, and $(b_1 \, , b_2) = (41/10, -19/6)$. 
Note that due to the $U(1)_Y$ and $SU(2)_W$ charges of  the SM quarks and leptons, the UED contributions to the $eeuu$ operator in Eq.~(\ref{eq:eeqq}) are always positive while for the $eedd$ operator, they are negative. From the experimental bounds of  Table~\ref{t:4fermibds} it is apparent that these two operators yield the strongest bounds on the UED parameter space.

\begin{table}[t]
\caption{Four Fermi contact interaction bounds in PDG(2010)  \cite{pdg}.}
\begin{center}
\begin{tabular}{c|c|c|c|c|c|c|c}
TeV&$eeee$& $ee\mu\mu$&  $ee\tau\tau$& $\ell\ell\ell\ell$&$qqqq$&$eeuu$&$eedd$ \\    
\hline
\hline
$\Lambda_{LL}^+$ &$>8.3$&$>8.5$&$>7.9$&$>9.1$&$>2.7$&$>23.3$&$>11.1$ \\
%\hline
$\Lambda_{LL}^-$  &$>10.3$&$>9.5$&$>7.2$&$>10.3$&$2.4$&$>12.5$&$>26.4$
\end{tabular}
\end{center}
\label{t:4fermibds}
\end{table}

\begin{figure}[t]%[htbp]
\centering
\centerline{ 
\epsfig{file=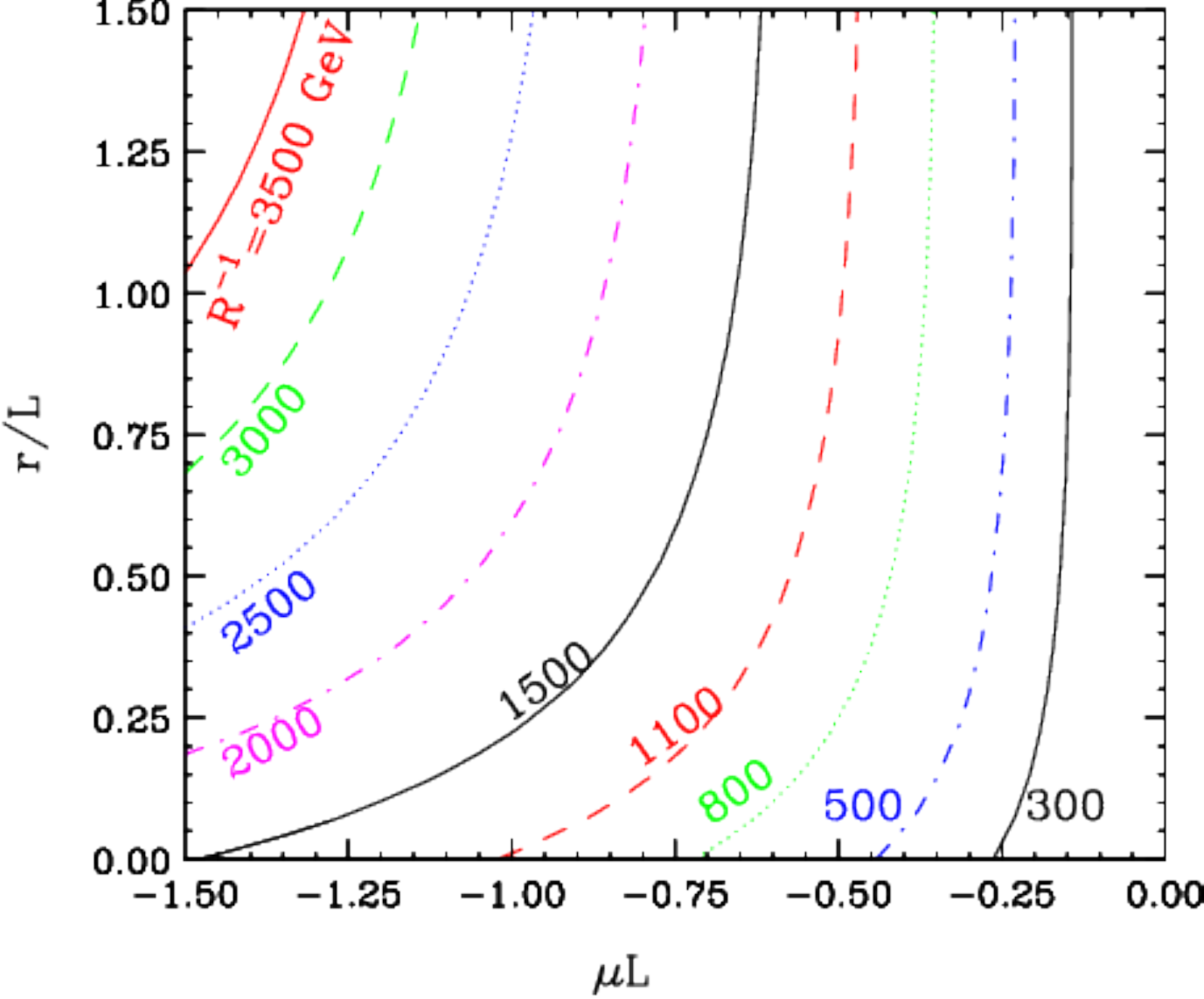, width=0.48\columnwidth} 
\hspace{0.2cm}
\epsfig{file=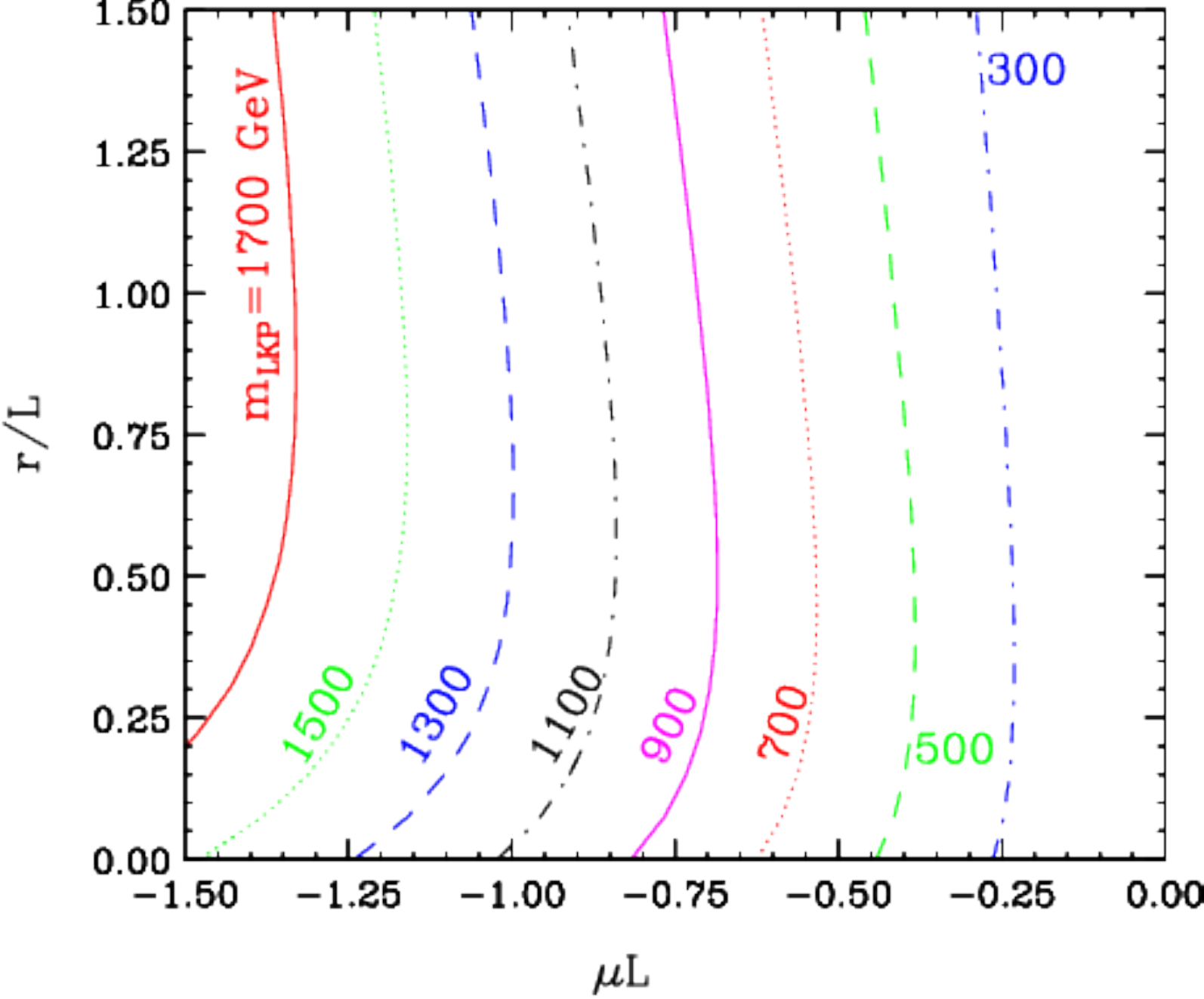, width=0.48\columnwidth} }
\caption{\sl Bounds on $R^{-1}$ from four-Fermi interactions. Contours represent minimally allowed values of $R^{-1}$ (left) and $m_{LKP}$ (right).}
\label{fig:4fermi} 
\end{figure}

In the left panel of Fig.~\ref{fig:4fermi}, we present contours of the lower bounds from the $eedd$ operator on $R^{-1}$ in the $r/L$ vs. $\mu L$ parameter space.\footnote{The $eeuu$ operator leads to slightly weaker bounds.} Like for the universal electroweak bounds, we translate these bounds into contours of the lower bounds on $m_{LKP}$ in the right panel of Fig.~\ref{fig:4fermi}.
Compared to the universal constraints in  Fig.~ \ref{fig:oblique}, the four fermion bounds are stronger for $\mu L\lesssim - 0.4$ while for small $\mu L$, the universal constraints dominate.

%%%%%%%%%%%%%%%%%%%%%%%%%%%%%%%%%%%%%%%%%%%%%%%%%%%%%%%
\subsection{Relic Abundance of KK Photon}
\label{sec:dm}
%%%%%%%%%%%%%%%%%%%%%%%%%%%%%%%%%%%%%%%%%%%%%%%%%%%%%%%

As shown in Section \ref{sec:ued}, both KK masses and couplings are modified significantly in the presence of bulk masses and brane localized terms, 
which in turn affects dark matter annihilations. 
KK fermion masses increase as the bulk mass $\mu$ increases (in the negative direction in our convention), 
while both KK fermion and boson masses decrease as the brane parameter $r$ increases.
Therefore there is an interesting interplay between two parameters regarding KK fermion masses but 
KK boson masses always get reduced. 
On the other hand, for a increasing $\mu$ in the positive direction, KK fermion becomes lighter than KK photon and may be a dark matter candidate.
For this reason, we restrict ourselves to $\mu < 0$, and consider the lightest KK photon as a potential dark matter candidate.

We follow the same procedure as in \cite{Huang:2012kz} and treat all KK masses differently and rescale $f_1$-$f_0$-$V_1$ coupling as in Eq.~(\ref{F110def})
($V_1$-$H_1$-$H_0$ and $V_1$-$V_1$-$H_0$-$H_0$ remain unchanged.).
The $f_1$-$f_0$-$V_1$ coupling gets reduced for an increasing $-\mu$, while its dependence on $r$ is weak, as shown in Fig. \ref{fig:couplings}.
In our calculation, we do not include coannihilation processes among KK fermions and KK photon, 
since generically there is relatively a large mass gap in the presence of a bulk mass and brane terms. 
Coannihilations become important only when the mass splitting is about 1\%, i.e., near the MUED limit. 
It is essentially the size of 1-loop radiative corrections in MUED, where the correction to masses of SU(2)W-singlet KK leptons is  $\sim$ 1\%.
Another important consideration would be resonant (co-)annihilation through 2-mode particles,  but for $\mu L\neq 0 \neq r/L$, the KK-level masses are not at multiples of $R^{-1}$ anymore, such that, unlike in MUED, resonant annihilation generically does not occur. See \cite{Belanger:2010yx} for effects of coannihilations and resonances in MUED.  

\begin{figure}[t]%[htbp]
\centering
\centerline{ 
\epsfig{file=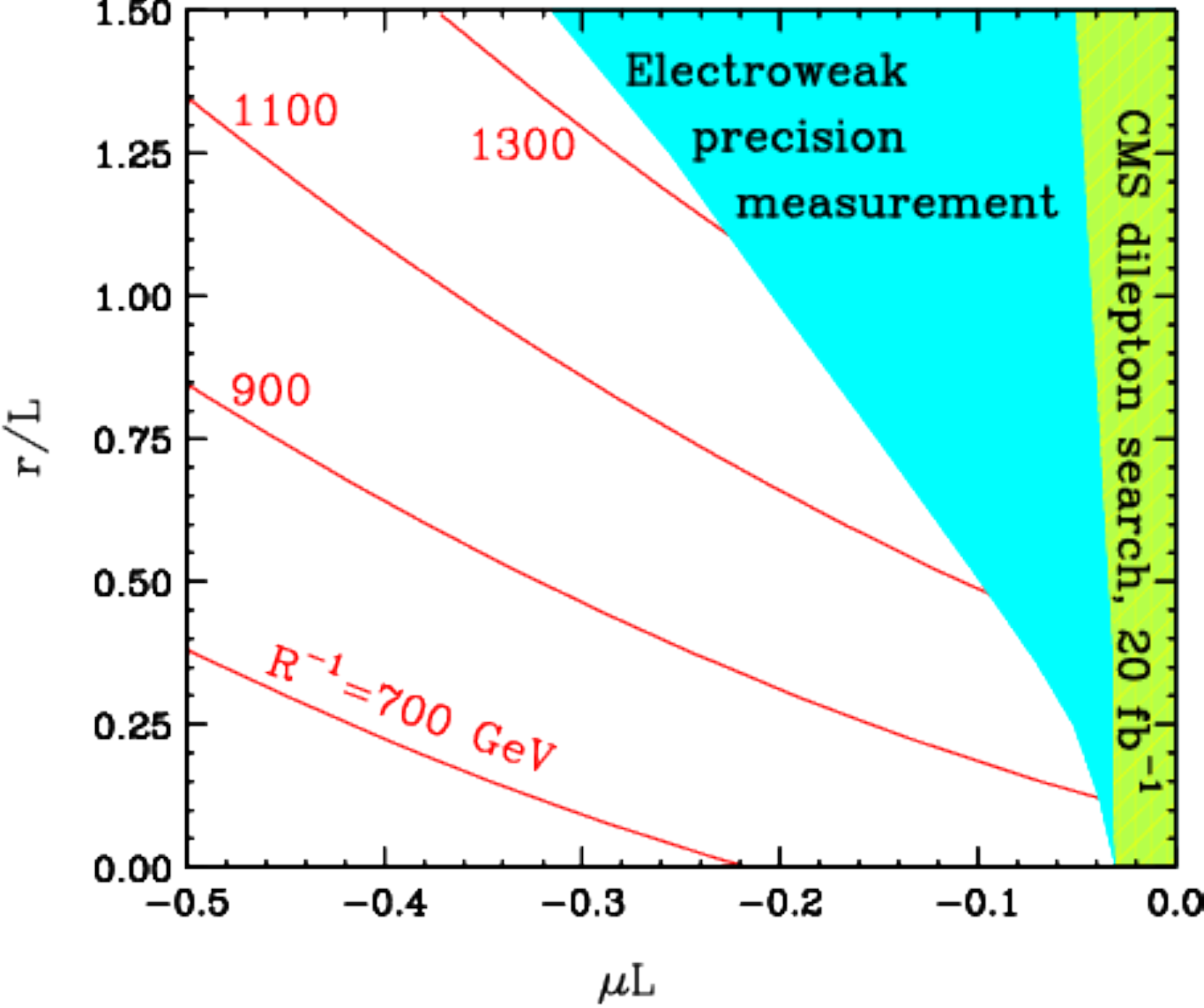, width=0.6\columnwidth} 
}
\caption{\sl Bounds from resonance search at the LHC and relic abundance of KK photon.}
\label{fig:colliderdm} 
\end{figure}

Our results are presented in Fig. \ref{fig:colliderdm}, where we show contours of $R^{-1}$, that is consistent with $\Omega h^2 = 0.1123$.
Each contour may serve as an upper bound on $R^{-1}$ for a given choice of $\mu$ and $r$, 
since KK photon could be one kind of dark matter species. 
However $R^{-1}$ greater than the corresponding value of the curve is not allowed since the model predicts too much dark matter. 
Combining with bounds from electroweak precision measurements and 4-Fermi interaction, which give lower bounds on $R^{-1}$, 
we obtain allowed region of ($\mu L$, $r/L$) space, shown in cyan in Fig. \ref{fig:colliderdm}. 
The yellow shaded region is allowed by the LHC  dilepton search as it will be discussed in the next section.
As shown in \cite{servanttait}, leptonic final states of KK photon annihilation are still dominant due to the nature of hypercharge interaction.

The pattern (slope of curves) shown in Fig. \ref{fig:colliderdm} can be understood easily based on KK masses and couplings.
In MUED ($r = \mu$), the WMAP consistent $R^{-1}$ is about 900 GeV.
An increasing bulk mass parameter lifts up KK fermion mass and reduces the $f_1$-$f_0$-$V_1$ coupling, which slows down efficient dark matter annihilation.
This results in lower values of $R^{-1}$. 
Now an increasing brane parameter reduces KK masses, leading to more efficient dark matter annihilation and hence increasing the $R^{-1}$ value.
For the $r=0$ limit, our result is consistent with that in \cite{Huang:2012kz}.

%%%%%%%%%%%%%%%%%%%%%%%%%%%%%%%%%%%%%%%%%%%%%%%%%%%%%%%
\subsection{Collider Phenomenology}
\label{sec:collider}
%%%%%%%%%%%%%%%%%%%%%%%%%%%%%%%%%%%%%%%%%%%%%%%%%%%%%%%

\subsubsection{Level-1 Modes}

Although we only determined the mass spectrum at tree-level, we can understand basic collider phenomenology from the modified mass spectrum and couplings.
In the entire parameter space ($\mu < 0$ and $r > 0$), level-1 KK fermions are always heavier than KK bosons, as shown in Fig. \ref{fig:masses}. 
Electroweak symmetry breaking lifts up $Z_1$ and $W_1^\pm$ slightly. 
Considering $R^{-1} \gg m_Z \, , m_W$, it is a minor correction ($\lsim {\cal O}(1)\%$), and radiative corrections are more important. 
Although they have not been calculated in this model context yet, we expect their effects to be similar to those in MUED, 
which are known to be approximately 1\% for $SU(2)_W$-singlet leptons, 3\% for $SU(2)_W$-doublet leptons, 
%\note{$SU(2)_W$-singlet/doublet leptons?} 
7\% for weak bosons, 20\% for KK quarks and 30\% for KK gluons, 
providing a fixed mass hierarchy, $m_{\gamma_1}~<~m_{\ell_1}~<~m_{W_1}~,~\,~m_{Z_1}~<~m_{Q_1}~<~m_{g_1}$.

In our model, the mass hierarchy is more complicated. 
First of all, it is likely that radiative corrections would still give the same hierarchy among gauge bosons, $m_{\gamma_1}<m_{W_1} , \, m_{Z_1} < m_{g_1}$.
The question is where the fermions reside.
In some parameter space (for a large $-\mu$ where $m_{Q_1} > m_{g_1}$), level-1 KK quarks decay to level-1 KK gluon in our model, $Q_1 \to g_1 q$, 
and level-1 KK gluons go through 3 body decays via virtual KK quarks, emitting two quarks, $g_1 \to q \bar{q'} V_1$ ($V=W\, ,~Z\, ,~\gamma$). 
Depending on the relative difference between $m_{Q_1} - m_{V_1}$ and $m_{Q_1} - m_{g_1}$ and couplings ($g_3$ vs. $g_2$), the decay $Q_1 \to q V_1$ may be open. 
Similarly, EW KK bosons would go through 3 body decays to the LKP ($\gamma_1$) plus two leptons.

Since the brane terms barely modifies $V_1$-$f_1$-$f_0$ but reduced KK masses, 
production cross sections of KK quark and KK gluons are enhanced for a large brane parameter for a given $\mu$, 
which generally reduces production cross sections due to heaviness of KK fermions and decreased $V_1$-$f_1$-$f_0$ coupling. 
As the phenomenology of level-1 KK modes depends largely on details of mass spectrum, we rather focus on level-2 KK modes. 

\subsubsection{Level-2 Modes: Resonance Searches at the LHC}

Level-2 modes in our model are very different from those in MUED. 
The mass hierarchy is repeated in higher KK modes in MUED but in our model, this is no longer true.
Without radiative corrections, for $r/L \ll 1$, $m_{f_2} \gsim m_{V_2}$ and for $r/L \gg 1$, $m_{f_2} < m_{V_2}$, 
while in MUED,  $m_{\gamma_2}~<~m_{\ell_2}~<~m_{W_2}~,~\,~m_{Z_2}~<~m_{Q_2}~<~m_{g_2}$. 
Among level-2 KK modes, KK bosons may appear as dijet or dilepton resonances, due to the non-zero $V_2$-$f_0$-$f_0$ coupling. 
For $-\mu L \gsim r/ L$, the decay of level-2 KK boson to two level-1 KK fermions is kinematically not accessible, 
while for $-\mu L \lesssim r/L$, the decay will become kinematically allowed, but suffers from phase space suppression. 
For a very small $-\mu L$, this could be compensated by the fact that the $V_2$-$f_0$-$f_0$ couplings are suppressed 
with respect to the $V_2$-$f_1$-$f_1$ couplings, but this would require $\mu L$ of an order at which we do not obtain bounds. 

In this section, we consider $\gamma_2$ and $Z_2$ resonances in the dilepton channel.
Their masses are given by Eq.~(\ref{gaugemass}) and decays of level-2 KK bosons to two SM particles are included in the decay widths. 
We use {\tt CalcHEP} to compute cross sections for $p\, p \to  \gamma_2+Z_2 \to \ell^+ \ell^-$ in the 2 dimensional parameter space ($\mu L$, $r/L$) for various values of $R^{-1}$.
Then we compare results with experimental data to set a limit on $R^{-1}$.
A search for narrow high-mass resonances decaying to electron or muon pairs has
been performed by both ATLAS and CMS experiments. 
In our study we follow the CMS analysis \cite{CMS:2012exa}, where upper limits have been set on the cross section times branching fraction for new boson production 
relative to the standard model Z boson production using 8 TeV data set.
The dimuon event sample used corresponds to an integrated luminosity of 20.6 fb$^{-1}$ 
while the dielectron event sample used corresponds to an integrated luminosity of 19.6 fb$^{-1}$.
Our results are shown in Fig. \ref{fig:colliderdm} where yellow shaded area is allowed by dilepton search at the LHC.
We used the maximally allowed $R^{-1}$ value for a given set of parameters to determine whether the point ($R^{-1}$, $r/L$, $\mu L$) is ruled out or not.
Hence the shaded region assumes that KK photon accounts for all of dark matter. 
However, smaller $R^{-1}$ for the same ($\mu L$, $r/L$) is ruled out, 
since KK resonances are lighter and will suffer from large signal cross sections. 
For low values of $r$, the limit from electroweak precision measurement is comparable to that of dilepton resonance search at the LHC with 8 TeV and 20 fb$^{-1}$.
Current bound on $\mu$ is $\mu L \lsim 0.05$ ($\mu \lsim 0.03/R$) and radiative correction to KK mass spectrum starts playing an important role. 
We find that the size of boundary term ($r$) is insensitive to resonance search and further investigation is needed. 

We have considered direct $s$-channel production of level-2 KK gauge bosons and their decays to the dilepton.
There are indirect contributions to the dilepton final state via decays of other level-2 KK particles. 
For instance, production of level-2 KK quarks and gluons via KK number conserving interaction is dominant at the LHC.
They decay to $Z_2$ and $\gamma_2$, which in turn appear as resonances. 
It is known that these indirect production may increase the LHC reach significantly \cite{2ndKKrefs} but require a complete knowledge of mass spectrum.
Our bounds should be interpreted as rather model-independent bounds from the direct production of level-2 KK gauge boson. 

We also looked at constraints from searches for dijet resonance and we found that the current bound obtained is weaker.
In the non-universal parameters, $\mu_L \neq \mu_Q$, two bounds (from dijet and dileptoon searches) should be considered separately.
Interestingly, the bulk mass is strongly constrained while the brane parameter remains free.

%%%%%%%%%%%%%%%%%%%%%%%%%%%%%%%%%%%%%%%%%%%%%%%%%%%%%%%
%%%%%%%%%%%%%%%%             Conclusions                   %%%%%%%%%%%%%%%%%%%%%%%%
%%%%%%%%%%%%%%%%%%%%%%%%%%%%%%%%%%%%%%%%%%%%%%%%%%%%%%%
\section{Conclusion}
\label{sec:conclusion}

In this paper we have generalized models with Universal Extra Dimensions in the presence of fermion-bulk masses and brane-localized terms.
Absence of FCNC leaves 19 free parameters, 13 of which reside in the fermion sector. 
For our detailed study, we introduced a universal brane term and a universal bulk mass, to avoid non-standard Higgs VEV and hybrid mixing between different KK modes.
We performed the KK decomposition and computed KK masses and couplings. 
Our results show that various experiments constrain the universal bulk mass strongly in its negative regime while the brane terms are relatively free. 

While in this paper we only concentrated on the universal parametrization, it should be kept
in mind that there are many interesting possibilities for extending the analysis to a more
general setup. For example, one could invoke different parameters for quark and lepton sectors. 
Collider and dark matter phenomenology of this extended model needs to be revisited in the full parameter space.

\bigskip
%%%%%%%%%%%%%%%%%%%%%%%%%%%%%%%%%%%%%%
%\section*{Acknowledgments}
\acknowledgments
%%%%%%%%%%%%%%%%%%%%%%%%%%%%%%%%%%%%%%
TF is supported by the National Research Foundation of Korea (NRF) grant funded by the Korea government (MEST) N01120547. 
He would also like to thank CERN for hospitality, where part of this work as been done.
KK is supported in part by the US DOE Grant DE-FG02-12ER41809 and by the University of Kansas General Research Fund allocation 2301566. 
SC is supported by Basic Science Research Program through the National Research Foundation of Korea 
funded by the Ministry of Education, Science and Technology (2011-0010294) and (2011-0029758).

%%%%%%%%%%%%%%%%%%%%%%%%%%%%%%%%%%%%%%%%%%%%%%%%%%%%%%%
%%%%%%%% %%%%%%%%                  Appendix         %%%%%%%%%%%%%%%%%%%%%%%%%%%
%%%%%%%%%%%%%%%%%%%%%%%%%%%%%%%%%%%%%%%%%%%%%%%%%%%%%%%

\appendix

%%%%%%%%%%%%%%%%%%%%%%%%%%%%%%%%%%%%%%%%%%%%%%%%%%%%%%%
%%%%%%%% %%%%%%%%                  Appendix   1      %%%%%%%%%%%%%%%%%%%%%%%%%%%
%%%%%%%%%%%%%%%%%%%%%%%%%%%%%%%%%%%%%%%%%%%%%%%%%%%%%%%

\section{Absence of Tree-Level FCNCs for Flavor-Blind Bulk Masses and Brane Terms}
\label{app:flavor}

The experimental bounds on FCNCs are a challenge to many BSM models. In our generalized UED model, the most general choices of bulk mass matrices $M_\Psi$, and boundary parameter matrices $r_\Psi$ and $r_{\lambda^{U,D,E}}$ will lead to tree level FCNCs and are thus excluded, unless $R^{-1}\gtrsim \mathcal{O} (10^3) \tev$.\footnote{See Ref.~\cite{Csaki:2010az} for a discussion within the split UED setup. Analogous arguments hold for UED with boundary terms.} 
In this appendix we show that no FCNCs  are present at tree level if one assumes $M_\Psi$, $r_\Psi$ and $r_{\lambda^{U,D,E}}$ to be flavor-blind, \ie proportional to the unit matrix in flavor space.

We have to show that at tree level, no flavor changing vertex of fermions with any KK mode of  neutral gauge bosons or the higgs exist. 
The derivation can be performed in analogy to the Standard Model. We start from the 5D action in Eq.~(\ref{5Daction}) and write the $SU(2)_W$ charged 5D fermions in the gauge eigenbasis as 
\beq
Q_i=\left(\begin{array}{c}
U_i^Q\\
D_i^Q
\end{array}\right)
\hspace{20pt} , \hspace{20pt}
L_i=\left(\begin{array}{c}
\nu_i^L\\
E_i^L
\end{array}\right),
\eeq
where $i$ is a flavor index. Expanding the 5D Higgs around its vacuum expectation value
\beq
H=\left(\begin{array}{c}
\chi^+(x,y)\\
\frac{1}{\sqrt{2}}\left(h(x,y)+v_5(y)+i\chi^3(x,y)\right) 
\end{array}
\right) \, ,
\label{Hvevexp}
\eeq
the Yukawa interactions yield an additional mass contribution
\bea
S_5\supset \int d^4 x \int dy &&\left\{\left(\delta_{ij}+r_{\lambda^E, ij}\left[\delta(y-L)+\delta(y+L)\right] \right)\frac{\lambda_{5,jk}^E v_5}{\sqrt{2}} \overline{E^L_i}E_k +\mbox{h.c.}\right.\nonumber\\
&+&\left.\left(\delta_{ij}+r_{\lambda^U, ij}\left[\delta(y-L)+\delta(y+L)\right] \right)\frac{\lambda_{5,jk}^U v_5}{\sqrt{2}} \overline{U^L_i }U_k +\mbox{h.c.}\right.\nonumber\\
&+&\left.\left(\delta_{ij}+r_{\lambda^D, ij}\left[\delta(y-L)+\delta(y+L)\right] \right)\frac{\lambda_{5,jk}^D v_5}{\sqrt{2}} \overline{D^L_i}D_k +\mbox{h.c.}\right\}\nonumber
\eea
in the 5D action. If $r_{\lambda^{U,D,E}}\propto \mathbbm{1}$, these mass terms can be diagonalized \emph{in the 5D action} by bi-unitary transformations. 
\bea
&E^L\rightarrow \tilde{E}^L=S^E E^L & E \rightarrow \tilde{E}=T_E E, \nonumber\\
&U^L\rightarrow \tilde{U}^L=S^U U^L & U \rightarrow \tilde{U}=T_U U, \nonumber\\
&D^L\rightarrow \tilde{D}^L=S^D D^L & D \rightarrow \tilde{D}=T_D D. \label{flavtrafos}
\eea
These transformations leave the fermion kinetic- and mass term invariant if all $r_{\Psi}$ and $M_\Psi$ are flavor-blind, {\it i.e.,} proportional to $\mathbbm{1}$. Therefore in this case, $\{\tilde{U}^L,\tilde{U},\tilde{D}^L,\tilde{D},\tilde{E}^L,\tilde{E}, \tilde{\nu}^L\}$ form a 5D mass eigenbasis. The interactions of these 5D fermions to 5D neutral gauge bosons and the Higgs are flavor diagonal. The only flavor violating interaction term to the 5D $W$ is in this basis is 
\beq
S_5\supset \int d^4 x \int dy \left\{\left(1+r_Q\left[\delta(y-L)+\delta(y+L)\right] \right)\left(i g V^5_{CKM,jk} \overline{\tilde{U}}_i^L W^+ \tilde{D}^L_k +\mbox{h.c.}\right)\right\},
\eeq
where $V^5_{CKM}=S_US^\dagger_D$ is the 5D analog of the CKM matrix. As in the standard model, we can use the fermion reparameterizations to choose $T_{U,D,E}=\mathbbm{1}=S_{U,E}$ and $S_D^\dagger=V^5_{CKM}$ with $\lambda^{U,D}_5$ diagonal with real, positive eigenvalues. As FCNCs are absent at the 5D level, the KK decomposition cannot lead to flavor violating interactions between any neutral gauge boson or Higgs KK modes and fermion KK modes. 

Note that in the proof above holds for a generic $y$-dependent VEV and also allows for differing bulk masses $M_\Psi$ and boundary parameters $r_\Psi$ and $r_\lambda^{U,D,E}$. Our choice of universal bulk masses and fermion boundary terms is not required for the absence of  tree level FCNCs and can be relaxed. It is only a simplifying assumption. In the following Appendices, we demonstrate how to obtain the KK mass spectrum and couplings. For universal bulk masses and boundary terms, the results are all given in terms of semi-analytical expressions, while for more general parameter choices, they have to be determined numerically. 
%Where appropriate, we comment on how to perform the generalized KK reduction.

%%%%%%%%%%%%%%%%%%%%%%%%%%%%%%%%%%%%%%%%%%%%%%%%%%%%%%%
%%%%%%%% %%%%%%%%                  Appendix   2      %%%%%%%%%%%%%%%%%%%%%%%%%%%
%%%%%%%%%%%%%%%%%%%%%%%%%%%%%%%%%%%%%%%%%%%%%%%%%%%%%%%

\section{Kaluza-Klein Decomposition}
\label{app:KKdecomp}
%\renewcommand{\theequation}{A.\arabic{equation}}
%\setcounter{equation}{0}
%%%%%%%%%%%%%%%%%%%%%%%%%%%%%%%%%%%%%%%%%%%%%%%%%%%%%%%

We derive the KK decomposition for fermions from the quadratic part of the Lagrangian in Eq. (\ref{FLag}) except for contributions from electroweak symmetry breaking which will be considered in Appendix \ref{app:masses}. For a 5D fermion with a left-handed zero mode the relevant part of the action reads\footnote{For a fermion with a right-handed zero mode, the KK decomposition is obtained with the replacements $L\leftrightarrow R$ and $\mu_\Psi\rightarrow -\mu_\Psi$.}
\bea
\SM_\Psi=\int_{-L}^L d^4x dy &&\left\{ \bar{\Psi}_L i\gamma^\mu  \overleftrightarrow{\partial_\mu}\Psi_L \left[1+r_\Psi \left(\delta(y+L)+\delta(y-L)\right)\right]+\bar{\Psi}_R i\gamma^\mu  \overleftrightarrow{\partial_\mu}\Psi_R\right.\nonumber\\
&&\left. -\bar{\Psi}_R\gamma_5 \overleftrightarrow{\partial_5} \Psi_L-\bar{\Psi}_L\gamma_5 \overleftrightarrow{\partial_5} \Psi_R-\mu_\Psi\theta(y)\bar{\Psi}_R\Psi_L-\mu_\Psi\theta(y)\bar{\Psi}_L\Psi_R
\right\},
\eea
where $\overline{\Psi} \overleftrightarrow{\partial_M} \Psi =\frac{1}{2}\{ \overline{\Psi} (\partial_M \Psi) - (\partial_M \overline{\Psi}) \Psi \} $.
Variation of the action yields the five dimensional equations of motion
\bea
[1+r_\Psi[\delta(y+L)+\delta(y-L)]]i\slashed{\partial} \Psi_L-(\partial_y+\mu_\Psi\theta(y))\Psi_R&=&0\label{EOM1} \, , \\
i\slashed{\partial} \Psi_R-(-\partial_y+\mu_\Psi\theta(y))\Psi_L&=&0\label{EOM2} \, .
\eea
Decomposing the five dimensional fermion field according to
\beq
\Psi (x,y)=\sum_{n=0}^\infty \left(\psi^{(n)}_L(x)f^{\Psi_L}_n(y)+\psi^{(n)}_R(x)f^{\Psi_R}_n(y)\right),
\eeq
the equations of motion, Eq.~(\ref{EOM1})-(\ref{EOM2}) are separated into the four-dimensional part
\beq
i\slashed{\partial}\psi^{(n)}_{L/R}=m_{f_n}\psi^{(n)}_{R/L}\label{EOM4D} \, ,
\eeq
and a $y$-dependent part
\bea
m_{f_n}[1+r_\Psi[\delta(y+L)+\delta(y-L)]f^{\Psi_L}_n-(\partial_y+\mu_\Psi\theta(y)) f^{\Psi_R}_n&=&0\label{EOM1a},\\
m_{f_n} f^{\Psi_R}_n-(-\partial_y+\mu_\Psi\theta(y)) f^{\Psi_L}_n&=&0\label{EOM2a}.
\eea
The bulk equations of motions are of identical to those in split UED \cite{sUED2}. 
To determine the boundary conditions on the branes, %we follow the procedure proposed in Ref.~\cite{Csaki1}. 
we first consider boundary kinetic terms located an $\epsilon$ distance away from the orbifold fixed points , \ie at $L-\epsilon$ and $-L+\epsilon$ and integrate the EOMs between $L$ and $L-2 \epsilon$ 
(or between $-L$ and $-L +2 \epsilon$)  whilst imposing Dirichlet boundary conditions  on the $Z_2$ parity odd $\Psi_R$ at $y=\pm L$. This yields
\beq
0=r_\Psi m_{f_n}  f^{\Psi_L}_n(L-\epsilon)-\left[ f^{\Psi_R}_n\right]^L_{L-2\epsilon}+\OM(\epsilon)=r_\Psi m_{f_n}  f^{\Psi_L}_n(L-\epsilon)+ f^{\Psi_R}_n(L-2\epsilon)+\OM(\epsilon),
\eeq
for the boundary term near $y=L$, and the analogous expression near $y=-L$. Taking the limit $\epsilon \rightarrow 0$, we obtain the effective boundary condition
\beq
r_\Psi m_{f_n}  f^{\Psi_L}_n(\pm L)=\mp f^{\Psi_R}_n(\pm L) \, .
\label{bdycondap}
\eeq

With the EOM Eqs.~(\ref{EOM1a})-(\ref{EOM2a}) and the boundary condition Eq.~(\ref{bdycondap}), the mass quantization condition for the KK modes and the wave functions can be determined up to their normalization. The normalization is then fixed by the modified orthogonality relations
 \bea
\int_{-L}^L dy \, f^{\Psi_L}_m f^{\Psi_L}_n[1+r_\Psi\left(\delta(y+L)+\delta(y-L)\right]=\delta_{mn},\nonumber\\
\int_{-L}^L dy \, f^{\Psi_R}_m f^{\Psi_R}_n=\delta_{mn}.
\label{fermionsclprdL}
\eea
 
%%%%%%%%%%%%%%%%%%%%%%%%%%%%%%%%%%%%%%%%%%%%%%%%%%%%%%%

\subsection{Zero Mode}
\label{sec:0mode}

A massless solution of Eqs.~(\ref{EOM1a})-(\ref{EOM2a}) is given by 
\bea
f^{\Psi_L}_0&=&\NM^\Psi_0 e^{\mu_\Psi |y|},\nonumber\\
f^{\Psi_R}_0&=&0\, .
\label{psizeromd}
\eea 
From Eq.~(\ref{fermionsclprd}) we find
\beq
\left|\NM^\Psi_0\right|^2= \frac{\mu_\Psi}{(1+2r_\Psi\mu_\Psi)e^{2\mu_\Psi L}-1},
\label{N0L}
\eeq
which is viable if
\beq
r_\Psi/L>\frac{e^{-2\mu_\Psi L}-1}{2\mu_\Psi L}, 
\label{ghostbound}
\eeq
where we expressed the constraint in terms of the dimensionless parameters $r/L$ and $\mu L$. Smaller $r/L$ lead to a negative (or vanishing) kinetic term of the zero mode, \ie a ghost (or non-normalizable) state, and are therefore excluded. In the ``light solutions'' in the next section, we will find an additional constraint on $r/L$.

 %%%%%%%%%%%%%%%%%%%%%%%%%%%%%%%%%%%%%%%%%%%%%%%%%%

\subsection{Kaluza-Klein Modes}
\label{sec:light}

A set of potential massive solutions  to Eqs.~(\ref{EOM1a})-(\ref{EOM2a}) consistent with the boundary conditions Eq.~(\ref{bdycondap}) is given by the ``light'' solutions
\bea
\mbox{for $n$ odd:}&&\left\{
\begin{array}{lll}
f^{\Psi_R}_n&=&\NM_n^\Psi \left(-\frac{k_n}{m_{f_n}}\cosh(k_n y)+\frac{\mu_\Psi}{m_{f_n}}\theta(y) \sinh(k_n y)\right),\\
f^{\Psi_L}_n&=&\NM_n^\Psi  \sinh(k_n y),
\end{array}\right.\\
\mbox{for $n$ even:}&&\left\{
\begin{array}{lll}
f^{\Psi_L}_n&=&\NM_n^\Psi  \left(\frac{k_n}{m_{f_n}} \cosh(k_n y)+\frac{\mu_\Psi}{m_{f_n}} \theta(y)\sinh(k_n y)\right),\\
f^{\Psi_R}_n&=&\NM_n^\Psi  \sinh(k_n y).
\end{array}
\right.
\eea
where
\beq
m^2_{f_n}=\mu_\Psi^2-k_n^2,
\label{mfrelnlight}
\eeq
and the wave numbers $k_n$ are determined by the mass quantization conditions
\bea
\mbox{ for $n$ odd:}&& k_n \cosh(k_n L)=(r_\Psi m^2_{f_n}+\mu)\sinh(k_n L),\label{mquantLlo}\\
\mbox{ for $n$ even:}&&r_\Psi k_n \cosh(k_n L)=-(1+r_\Psi \mu)\sinh(k_n L).\label{mquantLle}
\eea
The normalization factors are given by 
\beq
\NM^{\Psi}_n= \left\{
\begin{array}{lr}
\left (-L + \frac{\cosh (k_n L) \sinh(k_n L) }{k_n} + 2 r_\Psi \sinh^2 ( k_n L)\right ) ^{-1/2}&\mbox{ for $n$ odd,} \\
 \left (-L + \frac{\cosh (k_n L) \sinh(k_n L) }{k_n} \right )^{-1/2}&\mbox{for $n$ even.}\\
\end{array}
\right.
\eeq

If the mass quantization conditions allow for solutions of this type, these have to be included into the KK spectrum because the wave functions are required to form a complete function basis $\{f_n\}$, but as can be seen from Eq.~(\ref{mfrelnlight}), the solutions can lead to negative $m^2_{f_n}$, \ie a tachyon in the spectrum, which leads to an exclusion region in the parameter space.  

For negative values of $r_\Psi/L$ which do not lead to a ghost according to  Eq.~(\ref{ghostbound}), there exists a tachyonic which can be seen as follows: Eq.~(\ref{mquantLle}) can be rewritten as 
\beq
k_n L\coth(k_n L)=-\left(\frac{1}{r_\Psi/L}-\mu L\right) \, , 
\eeq
which for  $0 >r_\Psi/L > \frac{e^{-2\mu_\Psi L}-1}{2\mu_\Psi L}$ implies
\beq
\infty>k_n L\coth(k_n L)>\mu L \frac{1+e^{-2\mu L}}{1-e^{-2\mu L}}=\mu L \coth\left(\mu L\right).
\eeq
As $x\coth(x)$ is positive and monotonic for $x>0$ ($x<0$), this implies that a solution with $k_n^2>\mu^2$ exists. Therefore, the absence of ghosts and tachyons forbids $r_\Psi<0$.
\begin{figure}[t]%[htbp]
\centering
\centerline{ 
\epsfig{file=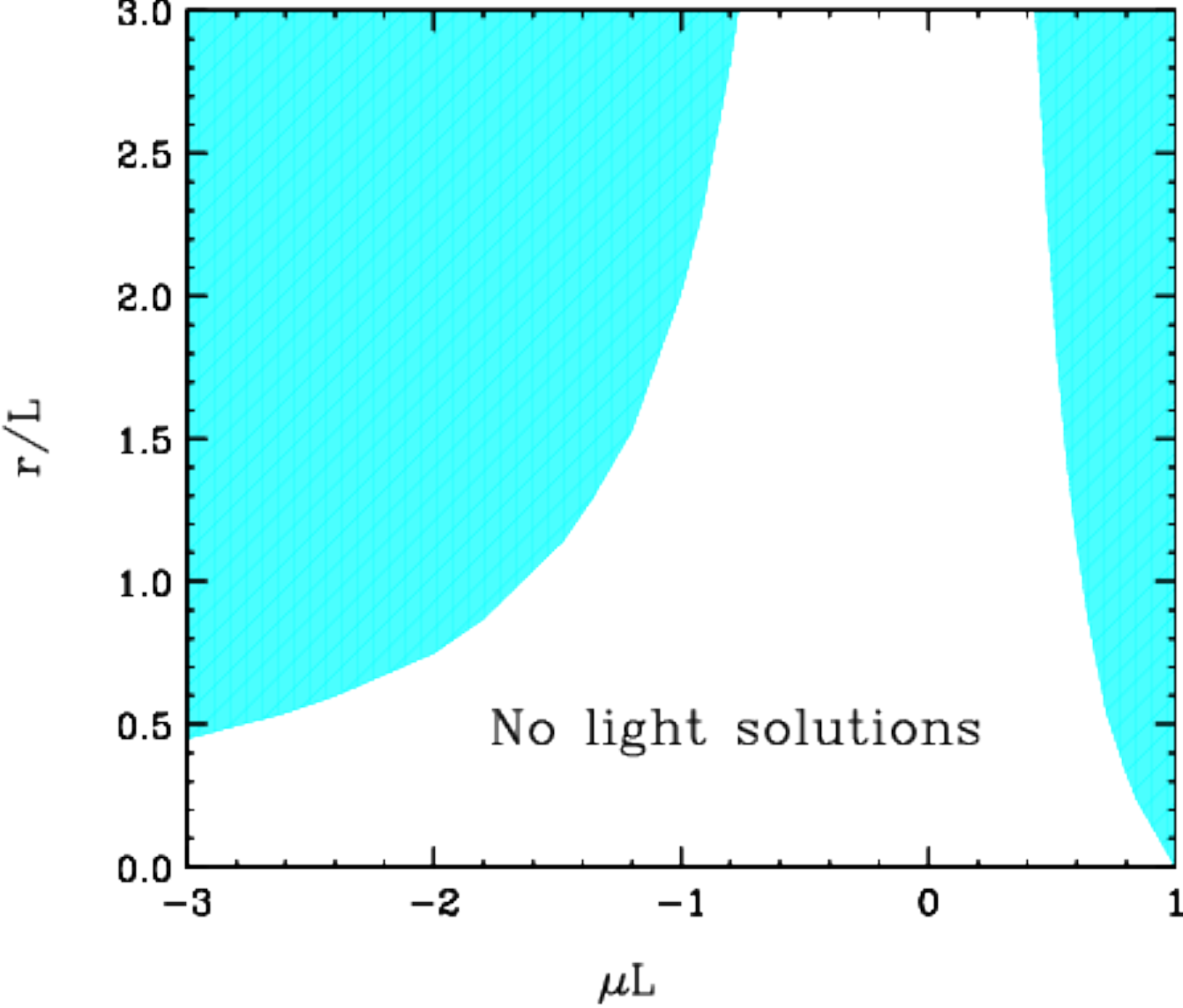, width=0.48\columnwidth} }
\caption{\sl Parameter space with and without a hyperbolic (``light'') first fermion KK mode.}
\label{fig:light} 
\end{figure}

For $r_\Psi>0$, we find that -- depending on $\mu L$ and $r_\Psi/L$ -- at most one solution to Eq.~(\ref{mquantLlo}) exists which, if present, is included in our analysis as the first KK mode in the spectrum. The parameter space with and without a light solution is shown in Fig.~\ref{fig:light}. We find no solution to Eq.~(\ref{mquantLle}) for $r_\Psi>0$. 

The remaining ``heavy'' solutions  to Eqs.~(\ref{EOM1a})-(\ref{EOM2a}) consistent with the boundary conditions Eq.~(\ref{bdycondap}) and 
their mass quantization condition have been given in Eqs.~(\ref{ffplus})-(\ref{Nf}), Section \ref{sec:KKdecomp}.
The KK decomposition of scalars and gauge bosons have already been discussed in Refs. \cite{BLKTrefs}. 
We in particular refer to Ref. \cite{FMP} for a detailed discussion of gauge fixing and electroweak symmetry breaking in the presence of BLKTs. 
The resulting wave functions and KK mass spectra are summarized in Eqs.(\ref{gaugeKKdecomp})-(\ref{gaugeKKmass}) in the main text.

\subsection{KK Decomposition for Fermions with a Right-Handed Zero Mode}\label{app:rhf}

The KK decomposition for 5D fermions $\Psi=U,D,E$ with a right-handed zero mode is performed analogously. The separated equations of motion are
\bea
i\slashed{\partial}\psi^{(n)}_{L/R}- m_{f_n}\psi^{(n)}_{R/L}&=&0 ,\\
m_{f_n}f^{\Psi_L}_n-(\partial_y+\mu_\Psi\theta(y)) f^{\Psi_R}_n&=&0\label{EOM1aR},\\
m_{f_n} [1+r_\Psi[\delta(y+L)+\delta(y-L)]f^{\Psi_R}_n-(-\partial_y+\mu_\Psi\theta(y)) f^{\Psi_L}_n&=&0\label{EOM2aR}.
\eea
The modified orthogonality relations are
\bea
\int_{-L}^L dy \, f^{\Psi_R}_m f^{\Psi_R}_n[1+r_\Psi\left(\delta(y+L)+\delta(y-L)\right]=\delta_{mn},\nonumber\\
\int_{-L}^L dy \, f^{\Psi_L}_m f^{\Psi_L}_n=\delta_{mn}.
\label{fermionsclprdR}
\eea
The resulting solutions read
\bea
n=0: && f^{\Psi_L}_0=\NM^{\Psi}_0 e^{-\mu_\Psi |y|},\\
\mbox{odd } n:&&\left\{\begin{array}{l}
f^{\Psi_R}_n=\NM^{\Psi}_n \sin(k_n y) \, , \label{ffplusR}\\
f^{\Psi_L}_n=\NM^{\Psi}_n \left(\frac{k_n}{m_{f_n}}\cos(k_n y)+\frac{\mu_\Psi}{m_{f_n}}\theta(y) \sin(k_n y)\right) \, ,
\end{array}
\right.\\
\mbox{even } n:&&\left\{\begin{array}{l}
f^{\Psi_R}_n=\NM^{\Psi}_n \left(-\frac{k_n}{m_{f_n}} \cos(k_n y)+\frac{\mu_\Psi}{m_{f_n}} \theta(y)\sin(k_n y)\right) \, ,\\
f^{\Psi_L}_n=\NM^{\Psi}_n \sin(k_n y) \, , \\
\end{array}
\right.
\eea
with
\beq
\begin{array}{ll}
 k_n \cos(k_n L)=(r_\Psi \left(m_{f_n}\right)^2-\mu_\Psi)\sin(k_n L)  \, , &  ~~{\rm for~ odd~ n}  \\
r_\Psi k_n \cos(k_n L)=-(1+r \mu_\Psi)\sin(k_n L) \, , & ~~  {\rm for~ even~ n}\,,
\label{fKKnumbR}
\end{array}
\eeq
and 
\beq
\NM^{\Psi}_n= \left\{
\begin{array}{lr}
\sqrt{  \frac{\mu_\Psi}{ ( 1 +2 r_\Psi \,  \mu_\Psi) \exp \left ( 2 \mu_\Psi L  \right ) - 1}} &\mbox{for } n=0\, ,\\
\frac{1}{\sqrt{L - \frac{\cos (k_n L) \sin(k_n L) }{k_n} + 2 r_\Psi \sin^2 ( k_n L) } }&\mbox{for odd } n\, ,\\
 \frac{1}{\sqrt{L - \frac{\cos (k_n L) \sin(k_n L) }{k_n}  } }&\mbox{for even } n.
\end{array}
\right.
\label{NfR}
\eeq
The ``light'' solutions are obtained from the heavy solutions by replacing $\sin / \cos$ with $\sinh / \cosh$. The parameter range in which ``light'' solutions are present can be read off from Fig.~\ref{fig:light} when replacing $\mu$ with $-\mu$.

The KK modes of 5D fermions with a right-handed zero mode resemble the KK mode decomposition with a left-handed zero mode when replacing $L\leftrightarrow R$ and $\mu\rightarrow -\mu$. With these replacements, the only difference is the relative sign between $f_n^{\Psi_L}$ and $f_n^{\Psi_R}$ at each non-zero KK level.

%%%%%%%%%%%%%%%%%%%%%%%%%%%%%%%%%%%%%%%%%%%%%%%%%%%%%%%%%%%%%%%
%%%%%%%%%%%%%%%                          Appendix 3        %%%%%%%%%%%%%%%%%%%%%%%%%%%%%%%%%
%%%%%%%%%%%%%%%%%%%%%%%%%%%%%%%%%%%%%%%%%%%%%%%%%%%%%%%%%%%%%%%

\section{Identification of the Underlying 5D Couplings}
\label{app:5Dto4D}

\subsection{Gauge Couplings}

In Section \ref{sec:mandcoupl}, we showed that the fermion-to-gauge boson zero mode coupling implies that the underlying 5D gauge couplings are related to the standard model couplings by $g^5_\AM=g_\AM\sqrt{2L(1+r/L)}$.
Here, we wish to discuss the generalization to the case $r_\Psi\neq r_\AM$ and demonstrate that the fermion-to-gauge boson coupling is consistent with the triple and quartic gauge boson couplings. For  $r_\Psi\neq r_\AM$, the  fermion-to-gauge boson zero mode coupling follows from\footnote{We assume a fermion with a left-handed zero mode, here. For a right-handed zero mode, the analogous calculation holds.}
\beq
\begin{split}
S_{eff}&\supset \int d^4x\, i g^5_\AM \overline{\psi}^{(0)} _{L/R}\slashed{\AM}^{(0)}\psi^{(0)}_{L/R} \int^L_{-L} dy f^\AM_0 f^{\Psi_{L}}_0f^{\Psi_{L}}_0 \left[1+r_\Psi\left(\delta(y+L)+\delta(y-L)\right)\right]\\
&=\int d^4x\, i g^5_\AM f^\AM_0 \overline{\psi}^{(0)} _{L/R}\slashed{\AM}^{(0)}\psi^{(0)}_{L/R} \,.
\end{split}\
\eeq
In the last step, we used the orthonormality relation Eq.~(\ref{fermionsclprd}) for fermions and the fact that $f^\AM_0$ is constant in $y$. 
The fermion-to-gauge boson zero mode coupling therefore implies
\beq
g^5_\AM=\frac{g_\AM}{f^\AM_0} \, .
\label{4D5Dgaugegen}
\eeq 
The triple and quartic gauge boson zero mode couplings follow analogously from the overlap integrals
\beq
g_\AM=\int dy  g^5_\AM \left(f^\AM_0\right)^3  \left[1+r_\AM \left(\delta(y+L)+\delta(y-L)\right)\right]=g^5_\AM f^\AM_0 \, , 
\eeq
and
\beq
g^2_\AM=\int dy  \left(g^5_\AM\right)^2 \left(f^\AM_0\right)^4  \left[1+r_\AM \left(\delta(y+L)+\delta(y-L)\right)\right]=\left(g^5_\AM f^\AM_0 \right)^2,
\eeq
and yield the same result Eq.~(\ref{4D5Dgaugegen}). We again used the flatness of $f^\AM_0$ and an orthonormality relation, Eq.~(\ref{gaugesclprd}) for the gauge bosons.

%%%%%%%%%%%%%%%%%%%%%%%%%%%%%%%%%%%%%%%%%%%%%%%%%%%%%%%%%%%%%%%

\subsection{Higgs Mass and Self-Interaction}

For the Higgs, we assume a boundary lagrangian
\beq
{\cal L}_{\partial H}=r_H\left(D_\mu H\right)^\dagger D^\mu H+  r_\mu \mu_5^2 |H|^2-r_\lambda \lambda_5 |H|^4 \, , 
\eeq
with $r_H=r_\mu=r_\lambda=r$. 
Minimizing the potential, we obtain 
\beq
v^5_{bdy}=\sqrt{2{\mu^2_5}{ \lambda_5}} =v^5_{bulk}\equiv v_5.
\label{vid}
\eeq
 Equality of the bulk and boundary VEV -- which resulted from choosing $r_\mu$ and $r_\lambda$ equal -- guarantee that the VEV of the system is independent of $y$. Expanding $H$ around $v_5$ according to Eq.~(\ref{Hvevexp}) furthermore yields
\beq
S_5 \, \supset \, \int d^4x dy \left[1+r_H\left(\delta(y+L)+\delta(y-L)\right)\right]\left(\frac{1}{2} \partial_M h \partial^M h-\mu^2_5 h^2\right) .
\label{higgsm}
\eeq
The equality of the boundary parameter of the kinetic and the mass term guarantee that the zero mode of the Higgs is flat and given by Eq.~(\ref{gaugezm}), 
while the KK mode wave functions are given by Eqs.(\ref{gaugeom})-(\ref{gaugeem}).\footnote{For differing boundary parameters, this expansion would hold for the kinetic term, but the mass term would induce KK mode mixing. 
Diagonalization of the KK mass matrix would then yield modified wave functions and would in particular imply a $y$ dependent zero mode.} 
One can find 
\beq
m_h=\sqrt{2}\mu_5=\sqrt{2}\mu_H \, ,
\label{Higgsmid}
\eeq
with the orthonormality relation in Eq.~(\ref{higgsm}), and 
\beq
\lambda_5=\lambda_H (2 L (1+\frac{r_H}{L})) \, ,
\label{Higgscid}
\eeq
from the KK decomposition of the 5D Higgs interaction terms, with flatness of $v_5$ and $f^H_0$. 

\subsection{Electroweak Sector}

Let us furthermore discuss the relation of the Higgs to the gauge boson masses in this generalized setup. 
We assume a universal boundary term in the electroweak sector, \ie $r_H=r_B=r_W$. 
This assumption simplifies the calculation of the gauge boson zero- and KK mode masses which in the effective action follows from
\bea
&&S_{eff}\supset \int d^4x \int_{-L}^L dy \left[1+r_H\left(\delta(y+L)+\delta(y-L)\right)\right]  \times \\
&&\hspace{1.5cm} \left\{\sum_{m,n} \left(-\frac{(g^5_Y)^2 v^2_5}{8}B^{(m)}_\mu B^{(n)\mu} f^B_mf^B_n
-\frac{g^5_Y g_w^5 v^2_5}{4}W^{3(m)}_\mu B^{(n)\mu} f^W_mf^B_n \right.\right.\\
&&\hspace{1.5cm}\left.\left.  -\frac{(g^5_w)^2 v^2_5}{8}W^{3(m)}_\mu W^{3(n)\mu} f^W_mf^W_n -\frac{(g^5_w)^2 v^2_5}{8}W^{+(m)}_\mu W^{-(n)\mu} f^W_mf^W_n
\right)\right\}.\nonumber
\eea
For $r_H=r_B=r_W$, the wave function bases $\{f^B_n\}$ and $\{f^W_n\}$ coincide, and are orthonormal with respect the scalar product depending on $r_H$, such that the above expression simplifies to 
\bea
S_{eff}\supset \int d^4x \left\{\sum_{n} \Big(\right.&&\left.\left.-\frac{(g^5_Y)^2 v^2_5}{8}B^{(n)}_\mu B^{(n)\mu} -\frac{g^5_Y g_w^5 v^2_5}{4}W^{3(n)}_\mu B^{(n)\mu} -\frac{(g^5_w)^2 v^2_5}{8}W^{3(n)}_\mu W^{3(n)\mu}\right.\right.\nonumber\\
&&\left.\left.-\frac{(g^5_w)^2 v^2_5}{8}W^{+(n)}_\mu W^{-(n)\mu}
\right)\right\}.
\eea
The zero mode of this expression together with Eq.~(\ref{4D5Dgaugegen}) allows us to identify 
\beq
v_5=\frac{v}{\sqrt{2L(1+r/L)}},
\label{v5ident}
\eeq
which is consistent with Eqs.~(\ref{Higgsmid})-(\ref{Higgscid}), and Eq.~(\ref{vid}).

The above discussion shows that our choice of an universal boundary parameter in the electroweak sector considerably simplifies the calculation and also guarantees that well-tested standard relations between electroweak parameters are satisfied. For a more general parameter choice, the 4D-to-5D parameter matching is more elaborate (see Ref.~\cite{FMP} for some initial work) and ultimately requires  performing a fit to the full set of electroweak data.

\subsection{Yukawa Sector} 

We discuss the down-type quark sector for illustration. We denote the 5D $SU(2)_W$ doublet quarks in the gauge eigenbasis as $Q_i=(U^Q_i,D^Q_i)^T$, where $i$ is a flavor index. 
As shown in Appendix \ref{app:flavor}, flavor-blind $r_\Psi$ and $M_\Psi$ allow to transform into the basis Eq.~(\ref{flavtrafos}) in which the 5D action is flavor diagonal. The 5D Yukawa couplings can be chosen diagonal and positive with the unitary transformation matrices given by $T_{U,D,E}=\mathbbm{1}=S_{D,E}$ and $S_D^\dagger=V^5_{CKM}$. In this basis, the Yukawa term in the 5D action Eq.~(\ref{5Daction}) yields a mass contribution
\bea
 S_{eff}&\supset \int d^4 x  &\left\{\sum_{m,n} \sum_i\left(\overline{\tilde{D}^{Q}}^{(m)}_{L,i} \frac{\lambda^D_{5,i} v_5}{\sqrt{2}} \tilde{D}^{(n)}_{R,i}+\mbox{h.c.}\right) \int_{-L}^L dy f^{Q_L}_m f^{D_R}_n \left[1+r_\lambda^D\left(\delta(y+L)+\delta(y-L)\right)\right]\right. \nonumber\\
 &&+\left.\sum_{m,n}\sum_i \left(\overline{\tilde{D}^Q}^{(m)}_{R,i}  \frac{\lambda^D_{5,i} v_5}{\sqrt{2}}  \tilde{D}^{(n)}_{L,i}+\mbox{h.c.}\right) \int_{-L}^L dy f^{Q,R}_mf^{D,L}_n \right\}\label{Yukfix} \, ,
 \eea
to the down-type quarks (and analogous for the up-type quarks and leptons). If $r_Q=r_D=r_\lambda^D$ and $M_Q=-M_D$, the functional bases $\{f^{Q,\pm}_n\}$ and $\{f^{D,\pm}_n\}$ coincide and are orthonormal with respect to Eq.~(\ref{fermionsclprd}), and we can directly identify
\beq
-\frac{\lambda^D_{5,i} v_5}{\sqrt{2}}=m_{d,i}=-\frac{\lambda^D_{i} v}{\sqrt{2}},
\eeq
which, together with Eq.~(\ref{v5ident}), implies $\lambda^D_{5,i}=\lambda^D_{i}\sqrt{2L(1+r/L)}$. For a more generic (flavor-blind) choice of fermion bulk- and boundary parameters, the Yukawa interactions induce flavor conserving KK mode mixing, which however is suppressed by $\sim m^2_{d,i}/m^2_{f_2}$ and thereby negligible. The identification of the underlying 5D Yukawa couplings can be calculated from the overlap integral of the zero modes in Eq.~(\ref{Yukfix}).

%%%%%%%%%%%%%%%%%%%%%%%%%%%%%%%%%%%%%%%%%%%%%%%%%%%%%%%
%%%%%%%% Appendix #4  %%%%%%%%%%%%%%%%%%%%%%%%%%%%%%%%%%%
%%%%%%%%%%%%%%%%%%%%%%%%%%%%%%%%%%%%%%%%%%%%%%%%%%%%%%%

\section{KK Fermion Mass Eigenbasis}
\label{app:masses}

We already discussed the relation of the gauge and mass eigenbasis of the electroweak gauge boson KK modes in Section \ref{sec:mandcoupl}, Eqs.~(\ref{gaugemeb})-(\ref{gaugemass}).
The mass eigenstates of the fermions follow from Eq.~(\ref{Yukfix}), which for our choice of parameters reduces to
\bea
S_{eff}\supset \int d^4x &&-\left\{\sum_{n=0} \left(\overline{\tilde{D}^{Q}}^{(m)}_{L,i} m_{d,i}\tilde{D}^{(n)}_{R,i}+\mbox{h.c.}\right)-\sum_{n=1}\left(\overline{\tilde{D}^{Q}}^{(m)}_{R,i} m_{d,i}\tilde{D}^{(n)}_{L,i}+\mbox{h.c.}\right)\right.\nonumber\\
&&\left.+\sum_{n=1}\left(m_{f_n}\overline{\tilde{D}^Q}^{(n)}_{L,i} D^{Q(n)}_{R,i}+m_{f_n}\overline{\tilde{D}}^{(n)}_{L,i} D^{(n)}_{R,i}\right)\right\},\label{fKKmix}
\eea
and analogous for the up-type quarks and leptons. Together with the field redefinition Eq.~(\ref{flavtrafos}), the zero mode mass eigenstates $u^{(0)}_i$, $d^{(0)}_i$, $e^{(0)}_i$ are related to the gauge eigenstates by  
\bea
u^{(0)}_{L,i}=U^{Q,(0)}_j & , & u^{(0)}_{R,i}=U^{(0)}_i,\\
d^{(0)}_{L,i}=V^\dagger_{CKM, ij}D^{Q,(0)}_j & , & d^{(0)}_{R,i}=D^{(0)}_i,\\
d^{(0)}_{L,i}=E{L,(0)}_j & , & e^{(0)}_{R,i}=E^{(0)}_i ,
\eea
and the masses given by the standard model masses. At each non-zero KK level in the down-sector, the four Weyl fermions $
\tilde{D}^{Q(n)}_{L,R}$ and $\tilde{D}^{(n)}_{L,R}$ mix to form two Dirac fermions $d^{(n)}_{1,2}$, which -- according to  Eq.~(\ref{fKKmix}) -- are
\bea
\left(\begin{array}{c}
d^{(n)}_{1 L/R,i}\\
d^{(n)}_{2 L/R,i}
\end{array}\right)&=&
\left(\begin{array}{cc}
 \cos(\alpha^{(n)}_i) & \pm \sin(\alpha^{(n)}_i)\\
\mp \sin(\alpha^{(n)}_i) &  \cos(\alpha^{(n)}_i) 
\end{array}\right)
\left(\begin{array}{c}
 \tilde{D}^{(n)}_{L/R,i}\\
 \tilde{D}^{Q(n)}_{L/R,i}
\end{array}\right)\nonumber\\
&=&
\left(\begin{array}{cc}
 \cos(\alpha^{(n)}_i) & \pm \sin(\alpha^{(n)}_i)\\
\mp \sin(\alpha^{(n)}_i) &  \cos(\alpha^{(n)}_i) 
\end{array}\right)
\left(\begin{array}{c}
 \delta{ij}D^{(n)}_{L/R,j}\\
 V^\dagger_{CKM,ij}D^{Q(n)}_{L/R,j}
\end{array}\right)\label{fmbasistrafo} \, ,
\eea
  where
  \beq
  \tan(2 \alpha^{(n)}_i)=\frac{m_{d,i}}{m_{f_n}},
  \eeq
 and masses
 \beq
 m_{d^n_{1,2 \, i}}=\sqrt{m^2_{f_n}+m_{d,i}}.
 \eeq
Leptons and the up-type quark sector can be treated analogous to the first line of Eq.~(\ref{fmbasistrafo}), \ie due to our choice of conventions, the field redefinitions do not depend on the CKM-Matrix.
For all fermionic KK excitations apart from the top KK modes, $m_{f,i}\ll m_{f_n}$, such that the mixing between the $SU(2)_W$ doublets and singlets in Eq.~(\ref{fmbasistrafo}) can be neglected.

\end{document}